\documentclass[fleqn,usenatbib]{mnras}

\usepackage{newtxtext,newtxmath}
\usepackage[T1]{fontenc}

\DeclareRobustCommand{\VAN}[3]{#2}
\let\VANthebibliography\thebibliography
\def\thebibliography{\DeclareRobustCommand{\VAN}[3]{##3}\VANthebibliography}

\usepackage{booktabs}

\usepackage{graphicx}	% Including figure files
\usepackage{amsmath}	% Advanced maths commands
\usepackage{lscape}

\title[Binning spectra in atmospheric retrievals]{The effect of spectroscopic binning on atmospheric retrievals}

\author[J. J. Davey et al.]{
Jack J. Davey,$^{1}$\thanks{E-mail: jack.davey.22@ucl.ac.uk (JJD)}
Kai Hou Yip,$^{1}$
Ahmed F. Al-Refaie$^{1}$
and Ingo P. Waldmann$^{1}$\\
$^{1}$Department of Physics and Astronomy, University College London, Gower Street, WC1E 6BT London, UK
}

\date{Accepted XXX. Received YYY; in original form ZZZ}

\pubyear{2023}

\begin{document}
\label{firstpage}
\pagerange{\pageref{firstpage}--\pageref{lastpage}}
\maketitle

% Abstract of the paper
\begin{abstract}
With the James Webb Space Telescope (JWST) offering higher resolution data in space-based transmission spectroscopy, understanding the capabilities of our current atmospheric retrieval pipelines is essential. These new data cover wider wavelength ranges and at much higher spectral resolution than previous instruments have been able to offer. Therefore, it is often appealing to bin spectra to fewer points, better constrained in their transit depth, before using them as inputs for atmospheric retrievals. However, little quantitative analysis of the trade-off between spectral resolution and signal-to-noise ratio has been conducted thus far. As such, we produce a simulation replicating the observations of WASP-39b by the NIRSpec PRISM instrument on board JWST and assess the accuracy and consistency of retrievals while varying resolution and the average photometric error. While this probes a specific case we also plot `binning paths' in the resulting sensitivity maps to demonstrate the best attainable atmospheric parameter estimations starting from the position of the real JWST Early Release Science observation. We repeat this analysis on three different simulation setups where each includes an opaque cloud layer at a different height in the atmosphere. We find that a much greater resolution is needed in the case of a high cloud deck since features are already heavily muted by the presence of the clouds. In the other two cases, there are large `safe zones' in the parameter space. If these maps can be generalised, binning paths could inform future observations on how to achieve the most accurate retrieval results. 

\end{abstract}

% Select between one and six entries from the list of approved keywords.
% Don't make up new ones.

%%%%%%%%%%%%%%%%%%%%%%%%%%%%%%%%%%%%%%%%%%%%%%%%%%

%%%%%%%%%%%%%%%%% BODY OF PAPER %%%%%%%%%%%%%%%%%%

\section{Introduction}

There is now substantial diversity in the population of known exoplanets and this is driving an ever increasing emphasis on characterising rather than just identifying exoplanets. In this pursuit, analysis of atmospheric transmission spectra has long been the most fruitful means of investigation. Through this technique, constraints have been placed not only on atmospheric compositions\,\citep{Barman2007_H2O,Tinetti2007_H2O,Swain2008_CH4,Pinhas2018_HotJupH2OSurvey}, but also on the temperature-pressure profiles and cloud properties\,\citep{Kreidberg2014_CloudsSuperEarth,1MacDonald2017_PatchyClouds_HD209} of many planets. However, from its inception to the the most recent observations, transmission spectroscopy has always been a field limited by the quality and quantity of data available.

Past spectra produced by the Hubble (HST) and Spitzer Space Telescopes (SST) were inherently limited by the signal-to-noise ratio (SNR) and spectral resolution that these instruments could achieve. As such, the chemical species that we could identify and the confidence with which we could constrain atmospheres was lower than necessary for detailed characterisation\,\citep{Swain2009_HD189733b_spec,Crouzet2012_XO2b_spec,Berta2012_GJ1214b_spec,Deming2013_HD209458b_spec,Swain2014_189733b_spec,Kreidberg2014_wasp43b_spec,Kreidberg2015_wasp12b_spec,Tsiaras2019_K218b_spec,Saba2022_wasp17b_spec}.

With the James Webb Space Telescope (JWST) now offering a vast improvement in the quality of data available for transmission spectroscopy, we are making great strides towards the goal of improved characterisation. Across several instruments and observing modes, JWST is able to offer transmission spectra covering a broader wavelength range and at a much higher spectral resolution than has previously been possible\,\citep{JWST_TransitingExoplanetSim,JWST_TransmissionSpecOverview}. 
Since with these new data we can study more atmospheres and in greater detail, it is possible to determine the atmospheric abundances of an ever increasing number of chemical species\,\citep[e.g. SO\textsubscript{2}, CO\textsubscript{2} and CH\textsubscript{4}][]{NIRSpecG395H_Wasp39b,JWST_NIRSpec_WASP39b_CO2,NIRCamWASP80b} on a more diverse sample of planets (such as Earth-sized planets studied theoretically by \cite{JWST_ExoplanetERSPlan}, \cite{JWST_ObservingEarthSizedExoplanets} and \cite{JWST_TransmissionSpecCapabilities} and shown in early releases of JWST data by \cite{JWST_EarthSizedExoplanSpec} among others). Additionally, with a larger population to analyse we can begin to employ statistics to make statements concerning general trends observed (as has already been attempted by \cite{Sing2016_TenHotJupiters}, \cite{Barstow2016_PopStudy}, \cite{HST_TsiarasPopStudy}, \cite{Fisher2018_PopStudy}, \cite{Pinhas2018_HotJupH2OSurvey}, \cite{changeat2022_PopStudy} and \cite{HST_EdwardsPopStudy} using the available HST and SST data and discussed generally by \cite{Bean2017_ComparativePlanetology}). Planned missions such as Ariel intend to exploit this large population of exoplanets and will focus specifically on common formation and evolution processes exhibited across the population\,\citep{ARIEL_Overview}. However, even in this pursuit, we are reliant on our ability to characterise individual planets first. Thus, with these new, more detailed spectra we face new challenges. We must now question how best to use these data in order to maximise the efficiency of our analysis.

Specifically, this investigation focuses on atmospheric retrieval frameworks. Retrievals offer the current, state of the art method in the analysis of transmission spectra. With the concept first being discussed by\,\cite{FirstRetrievals}, many retrieval codes have now been developed (e.g. NEMESIS\,\citep{NEMESIS}, CHIMERA\,\citep{CHIMERA}, TauREx\,\citep{TauREx1,TauREx2,TauREx3} and petitRADTRANS\,\citep{petitRADTRANS}, with a more comprehensive list presented in\,\cite{ListRetrievalCodes_2023}). Whilst HST and SST were marred by comparatively poor spectral resolution and high uncertainties, JWST offers unprecedented observational data quality. However, dealing with these large, complex data sets can present issues of their own. 

It is imperative that we understand the limits of results attainable from studies of atmospheres on individual exoplanets and the remaining obstacles with our current methods of analysis.  Studies of temporal binning during the lightcurve fitting procedure have been carried out for both transit\,\citep{TemporalBinning_Transit} and phase curve observations\,\citep{TemporalBinning_PhaseCurves}. Additionally, studies of the sensitivity of high resolution spectroscopy using cross correlation techniques to probe exoplanet atmospheres have recently been conducted by \cite{HRS_SensitivityStudy}. However, to date, there has been little attention given to binning data at the spectrum level (i.e. binning in wavelength space) for use with medium and low resolution spectra observed with space-based telescopes.

In this work, we evaluate the trade-off between spectral resolution and the scale of photometric error when binning spectra. This is quantified by considering the accuracy of parameter estimations made in atmospheric retrievals on data binned to varying resolutions. When binning spectra, we must consider three separate but related effects:
\begin{enumerate}
    \item The achievable uncertainty on the retrieved parameters
    \item The induced bias on the retrieved values of each parameter
    \item The change in correlations observed in the conditional posterior distributions
\end{enumerate}

Given that a much lower SNR will be recovered by observations of smaller planets, understanding the effectiveness of binning as a way of improving the SNR of data by sacrificing spectral information will be a crucial step towards probing this part of the exoplanet population. In an ideal scenario, given that we know the observing resolution that facilities such as JWST and Ariel offer, this type of research could advise on how long to observe a given target in order to reach the photometric error which would maximise the accuracy of its parameter predictions, possibly with the aid of subsequent spectral binning. This would require a generalised framework but in this initial investigation we use the specific example of planets similar to WASP-39b\,\citep{Wasp39b_Discovery}, a hot, Saturn-sized planet observed with JWST\,\citep{NIRSpecG395H_Wasp39b,JWST_NIRSpec_WASP39b_CO2,NIRSpecPRISM_Wasp39b,NIRCam_WASP39b,NIRISS_WASP39b}.

We wish to determine the optimal binning regime which preserves sufficient information for a retrieval while also reducing the computational demands of the forward modelling procedure. Similarly, by investigating the trade-off of higher-SNR spectral points at reduced spectral resolution, we can hope to identify optimal observing strategies for JWST.
Through simulations, we analyse WASP-39b across the wavelength range of the NIRSpec PRISM instrument~\citep{NIRSpecOverview} in three distinct cases; one with a cloud-free atmosphere, one with low altitude clouds and one with high altitude clouds. By doing so, we are able to explore the importance of the height of spectral features in a spectrum and comment on how it affects the previous three considerations. (Additional details on WASP-39b can be found in section~\ref{sec:WASP39b} and details concerning the simulations in section~\ref{sec:Methodology}.)

We investigate the dependency of these effects on resolution and error using two independent methods. The first is by generating simulated spectra across a grid of resolutions and errors. These spectra have constant resolution across the wavelength range of the NIRSpec PRISM instrument and the error bars on each point are equal. While this is a simplistic assumption, it allows us to accurately position each within a resolution-error grid and study the change across this parameter space.

The second method uses a more realistic binning scheme which retains some of the variation across the instrument. We start from the wavelength grid of the original WASP-39b observation as part of the JWST Early Release Science (ERS) program\,\citep{NIRSpecPRISM_Wasp39b} and, propagating the errors reported, we bin down to lower resolutions. (The method by which this is achieved is described in section~\ref{sec:AdditionalInvest}.)

In this investigation, we expect to be able to map the resolution-error space and mark boundaries between poorly retrieved regimes and `safe zones'. In these safe zones, binning will not significantly alter the retrieval results but other regions (with larger errors or lower resolution) may see rapid changes in the accuracy of retrieval predictions. Given the non-linear nature of the effects of binning within the retrieval process, these safe zones must be mapped out numerically as described above.

\subsection{Motivation for binning spectra}
\label{sec:BinMotivation}

Atmospheric retrievals employ forward modelling and Bayesian sampling processes to find the set of input parameters which best replicate an observed spectrum~\citep{RetrievalSummaryChap}. This is accomplished through the process of drawing a set of parameters from some pre-defined prior space, computing a forward model using equations of radiative transfer and, then, evaluating a likelihood function to quantify the level of agreement between the observed and simulated data. The process then repeats until some limit or level of accuracy is achieved.

Many factors affect the efficiency of this process. The complexity of the forward model calculations, the number of parameters for which the retrieval is fitting and the size of the data set are just a few but already provide plenty of scope for investigations. When looking to increase efficiency, it is always important to bear in mind any consequent sacrifice in the accuracy of the output estimations. 

Binning spectra in wavelength space can be a useful method when trying to extract the maximum possible information from a spectrum while also giving consideration to the computational cost of retrievals or other fitting procedures. However, the consequences of binning can be severe. It is a process which results in an inevitable loss of information. As such, a balance should be struck between the reduction in the photometric error by binning an observation and the loss of morphological information as spectral resolution decreases.

Spectral resolution ($R$) can be calculated via the equation
\begin{equation}
    R = \frac{\lambda}{\Delta\lambda}
\end{equation}
\noindent which is the ratio of the observed wavelength ($\lambda$) and the minimum separation to the next observable feature ($\Delta\lambda$). As such, a higher resolution spectrum will have more points than a lower resolution spectrum across the same wavelength range. All instruments will have some intrinsic maximum resolution which is set by the dispersion element and detector characteristics. However, binning down spectra remains an option later in the data processing pipeline.

When a set of spectral points are binned together, their combined photon counts contribute to the strength of the signal. Given that this is a counting process, Poisson statistics apply and the result is a reduction in photometric error proportional to $1/\sqrt{\text{N\textsubscript{photons}}}$ (where N\textsubscript{photons} is the number of photons contributing to the signal). Thus, the resulting set of spectral points can be better localised in their transit depth but at the expense of being less well constrained in their wavelength.

In our analysis we compute approximate `binning paths' to give an idea of how a specific dataset can move in the resolution-error space under investigation with additional binning. These paths are calculated based on the number of bins remaining in our simulated spectra and the corresponding reduction in photometric error assuming, naively, that there are equal photon counts in every bin in a spectrum. The errors are then reduced by a factor
\begin{equation}
    \sigma_{2} = \sigma_{1} \left( \frac{\sqrt{N_{Bins}(R_{2})}}{\sqrt{N_{Bins}(R_{1})}} \right)
    \label{Eq:BinningPath}
\end{equation}
\noindent where the $\sigma_{i}$ refer to the errors on the spectral bins at their corresponding resolutions of $R_{i}$, and $N_{Bins}(R_{i})$ refers to the number of bins (or spectral points) at a resolution of $R_{i}$. These paths should not be seen as definitive. Instead, they should be used as guides to indicate the resolution-error combination possible given a specific start point in the parameter space.

\begin{table}
    \centering
    \begin{tabular}{c|c|c}%|c}
    \toprule\toprule
         & \textbf{Parameter} & \textbf{Value} \\%& \textbf{Ref}\\
         \midrule
         \textbf{Planetary}  & Mass (M\textsubscript{Jup}) & 0.28\textsuperscript{a} \\%& \cite{StellarParameters_Bonomo}\\
         Values from         & Radius (R\textsubscript{Jup}) & 1.27\textsuperscript{a} \\%& \cite{StellarParameters_Bonomo}\\
         \cite{StellarParameters_Bonomo}\textsuperscript{a} & Effective Temperature (K) & 1000\\
         and \cite{WASP39bParameters_Maciejewski}\textsuperscript{b} & Impact Parameter & 0.45\textsuperscript{b} \\%& \cite{WASP39bParameters_Maciejewski}\\
                             & Orbital Period (days) & 4.05\textsuperscript{a} \\%& \cite{StellarParameters_Bonomo}\\
        \midrule
        \textbf{Stellar}    & Mass (M\textsubscript{\(\odot\)}) & 0.93 \\%& \cite{StellarParameters_Bonomo}\\
        All values from     & Radius (R\textsubscript{\(\odot\)}) & 0.895 \\%& \cite{StellarParameters_Bonomo}\\
        \cite{StellarParameters_Bonomo} & Temperature (K) & 5400 \\%& \cite{StellarParameters_Bonomo}\\
        \midrule
        \textbf{Chemical}              & H\textsubscript{2}O & -3.5\\
        \textbf{abundances}            & CO\textsubscript{2} & -5.5\\
        (\textit{All in log scale})    & CO & -3.5\\
                                       & SO\textsubscript{2} & -6\\
                                       & Na & -5\\
        \midrule
        \textbf{Fill gases}            & H\textsubscript{2} & 0.17\\
        (\textit{Ratio of remaining}   & He & 1\\
        \textit{atmosphere}) \\
        \midrule
        \textbf{Atmospheric layers} & High clouds (Pa) & 10\\
        (\textit{3 cloud cases})    & Low clouds (Pa) & 1000\\
                                    & No clouds (Pa) & N/A\\
                                    & Min. pressure level (Pa) & 10\textsuperscript{-1}\\
                                    & Max. pressure level (Pa) & 10\textsuperscript{6}\\
        
    \end{tabular}
    \caption{The inputs parameters used to generate the simulation spectrum. This simulation is designed to replicate the shape of the spectrum observed by the NIRSpec PRISM on JWST for the planet WASP-39b. A base spectrum was generated using parameters from the relevant literature but the chemical abundances were altered to best match the reference observations.}
    \label{table:Sim}
\end{table}

\begin{table}
    \centering
    \begin{tabular}{c|c|c}
    \toprule\toprule
         & \textbf{Name} & \textbf{Opacity reference} \\
         \midrule
         \textbf{Atomic and molecular} & H\textsubscript{2}O & \cite{H2O}\\
         \textbf{cross-sections} & CO\textsubscript{2} & \cite{CO2}\\
                                 & CO & \cite{CO}\\
                                 & SO\textsubscript{2} & \cite{SO2}\\
                                 & Na & \cite{Na_K}\\
         \midrule
         \textbf{Collisionaly induced} & H\textsubscript{2}-H\textsubscript{2} & \cite{H2_CIA_HighTemp}\\
         \textbf{absorption}           & H\textsubscript{2}-He & \cite{H2_CIA_LowTemp}\\
         &&\cite{RICHARD2012_HITRAN_CIA}\\
         &&\cite{CIA_Hitran_2013}
    \end{tabular}
    \caption{A list of opacity sources used for the generation of forward models in this investigation. All sources come from the ExoMol project \protect\citep{ExoMol1,ExoMol2,ExoMol3} or the HITRAN database \protect\citep{RICHARD2012_HITRAN_CIA,CIA_Hitran_2013}.}
    \label{table:Opac}
\end{table}

\begin{figure*}
    \includegraphics[width=\textwidth]{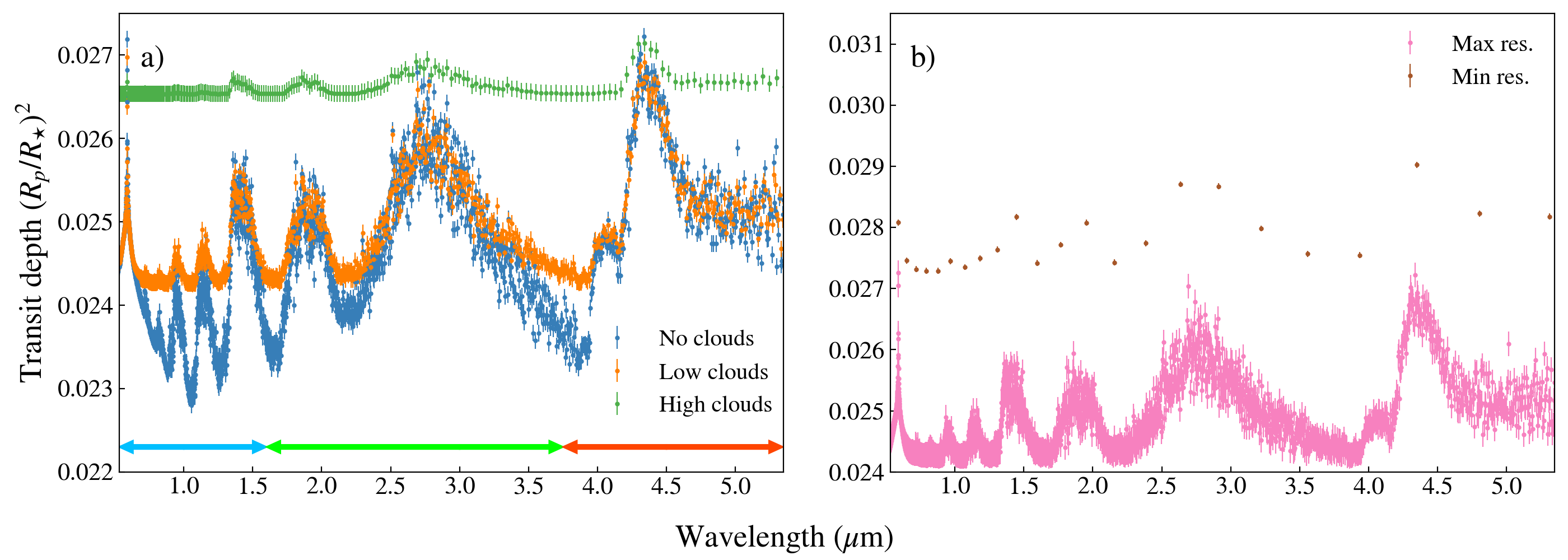}
    \caption{Simulated spectra for the atmosphere of WASP-39b based on the JWST observations presented in~\protect\cite{NIRSpecPRISM_Wasp39b} using the FIREFLy data reduction pipeline (\protect\cite{FIREFLy}). In panel \textbf{a)} the three sets of points represent different instances of the assumed cloud deck pressure and spectral resolution. In each case the error bars are constant across the wavelength range and have been set to 100\,ppm in this figure to reflect the average error quoted on the WASP-39b NIRSpec PRISM spectrum. In blue is the cloud-free case at a resolution of R~=~900; shown in orange is the case with low level clouds (at a pressure of 1000\,Pa) and R~=~500; in green is the case with high clouds (at a pressure of 10\,Pa) and R~=~100. The three arrows (in blue, green and red) show the bins used when calculating the average error in particular wavelength ranges. The limits of these regions are chosen arbitrarily but, broadly speaking, align with prominent features in the spectrum. In \textbf{b)}, spectra demonstrating the extreme cases of both binning and photometric error considered in this investigation are plotted. In pink is a spectrum using the highest resolution, R~=~1100 and with the largest error, $\sigma$~=~200\,ppm, while in red, the resolution is R~=~10 and $\sigma$~=~50\,ppm (the minimum values). Note that both spectra in panel \textbf{b)} use the low cloud case and the spectrum shown in black has been shifted by $(R_{p}/R_{\star})^{2}$~=~0.003 for visibility.}
    \label{fig:Simulations}
\end{figure*}

\subsection{WASP-39b}
\label{sec:WASP39b}

Due to its large scale height and relatively inactive host star, the hot Saturn exoplanet, WASP-39b\,\citep{Wasp39b_Discovery}, is an ideal candidate for transmission spectroscopy studies. It has been observed with multiple instruments using HST and SST\,\citep{HST_STIS_Wasp39b,HST_WFC3_WASP39b}, the Very Large Telescope\,\citep[VLT][]{VLT_WASP39b}, the William Herschel Telescope\,\citep{LRG_BEASTS_WASP39b} and, more recently, with JWST\,\citep{NIRSpecG395H_Wasp39b,NIRSpecPRISM_Wasp39b,NIRCam_WASP39b,NIRISS_WASP39b}. In combination, these observations provide wavelength coverage of approximately 0.3\,\textmu m to 5.5\,\textmu m.

These observations have given a good deal of insight into the composition and structure of WASP-39b's atmosphere but degeneracies and discrepancies within and between these analyses remain. In particular, there is disagreement in the quoted metallicity and corresponding dependence on the presence of clouds and/or hazes in the atmosphere\,\citep{HST_STIS_Wasp39b,LRG_BEASTS_WASP39b,arfaux2023_CloudsAndHaze_WASP39b}. Since both of these physical features can serve to reduce the observed transit depth by suppressing the observable scale height in transmission, they often lead to conflicting results in the literature. WASP-39b was also found to be the first example of an exoplanet where photochemistry is observed in the atmosphere\,\citep{photochem_wasp39b} and, as such, it is a valuable target and a good testing ground for the capabilities of the JWST NIRSpec instrument\,\citep{NIRSpecOverview} in observing atmospheres of exoplanets in transit.

Given that there are multiple studies of this system, full details of the system parameters used for these investigations are shown in the `\textbf{Planetary}' and `\textbf{Stellar}' sections of table~\ref{table:Sim} along with their appropriate references. While this investigation will focus on the specific case of planets similar to WASP-39b, future studies will consider how easily these results could be generalised and applied to other planets (of the same or differing planetary type) and other instruments.

\section{Simulation and Retrieval Methodologies}
\label{sec:Methodology}

For the purposes of this investigation, simulations were created to mimic the form of the JWST Early Release Science (ERS) observations of WASP-39b over the wavelength range of the NIRSpec PRISM instrument\,\citep{NIRSpecPRISM_Wasp39b}. Multiple spectra are generated since we vary the spectral resolution (between 10 and 1000) and the photometric error on each data point (between 50\,ppm and 200\,ppm).

These spectra are then input as observations to the TauREx\,3 retrieval code\,\citep{TauREx1,TauREx2,TauREx3} and the resultant parameter estimations are compared to the known values. We quantify accuracy through the percentage deviation of the predictions compared to the know input values using the equation,
\begin{equation}
    Deviation (\%) = \frac{\vert \theta_{fit}-\theta_{true}\vert}{\vert \theta_{true}\vert}\times 100,
\end{equation}
\noindent where $\theta_{fit}$ is the retrieved value and $\theta_{true}$ the value input to the simulation for the parameter, $\theta$.

We then plot `bias maps' using the deviation value. Doing so, we can study the region (in resolution-error space) where the actual observations lie and determine how well they can be retrieved at their native resolution and at binned resolutions. To investigate the importance of spectral feature height in the retrievals, spectra with different cloud decks have been considered and are investigated using identical retrieval and analysis methods on each (described in detail in the sections \ref{sec:SimulationSetUp} and \ref{sec:RetrievalSetUp}).

In addition to this basic investigation, we also give consideration to other possible effects which could bias our resulting sensitivity maps. Specifically, we look into model parameterisation, the influence of our choice of wavelength grid and the binning strategy used to reduce the resolution. 

\subsection{Simulation set up}
\label{sec:SimulationSetUp}

\begin{figure}
    \includegraphics[width=\columnwidth]{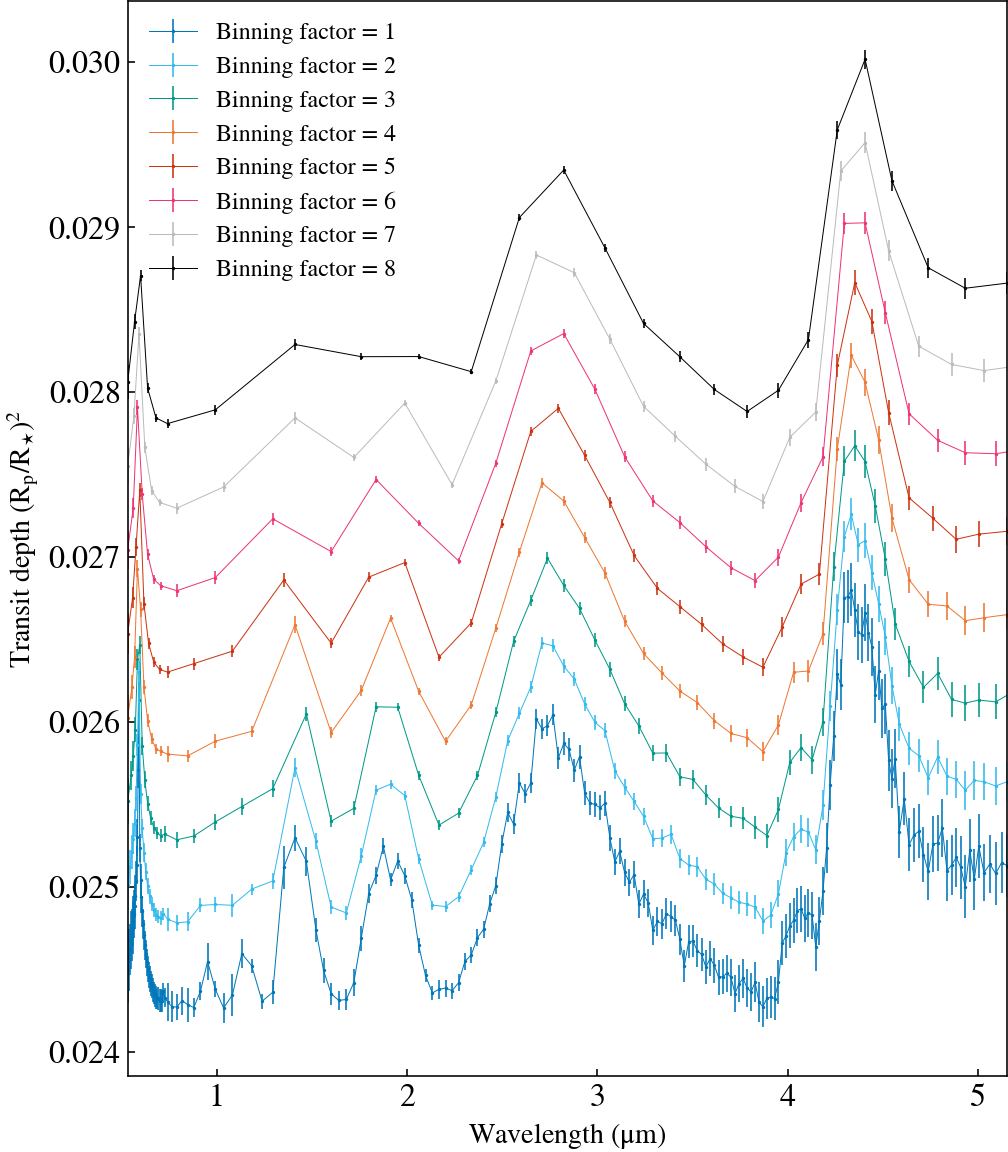}
    \caption{The different spectra obtained by binning the simulation of a planet like WASP-39b to the wavelength grid of the JWST NIRSpec PRISM data from the ERS program \citep{NIRSpecPRISM_Wasp39b} and then reducing the number of points in the spectrum before binning further. All spectra are produced by TauREx and propagate the error bars from the ERS data to the new wavelength grid. This method preserves more of the variation in resolution and error scale across the wavelength coverage of the instrument than binning to a constant resolution grid with constant error bars. Each spectrum receives an artificial offset in transit depth in this plot so that they can be distinguished from one another.}
    \label{fig:BinFactorSpec}
\end{figure}

In order to study the deviation of retrieved values from the `true' value, simulated data sets were generated so that the specific input parameters were known. Simulations were generated for three different cloud scenarios. These included the case of no clouds being present in the atmosphere, a high altitude (low pressure, 10\,Pa) grey cloud deck and a low altitude (high pressure, 1000\,Pa) grey cloud deck. The cloud level is the only parameter which differs between the three forward models, every other parameter was fixed to a constant value.

The simulations used a typical set up for a hot gaseous planet. Excess atmospheric content was filled with H\textsubscript{2} and He in the ratio [0.17:1] once dominant gas abundances had been specified. Molecular contributions were included due to H\textsubscript{2}O, CO, CO\textsubscript{2} and SO\textsubscript{2}. Na was also included as an atomic feature at the short wavelength end of the spectrum. Contributions due to collisionally induced absorption (C.I.A.) of H\textsubscript{2}-H\textsubscript{2} and H\textsubscript{2}-He pairs are also included. All opacities were taken from the ExoMol project \citep{ExoMol1,ExoMol2,ExoMol3}. Table~\ref{table:Opac} provides the appropriate references for each opacity calculation.

We use an isothermal temperature profile through an atmosphere of 100 layers with the pressure varying between 10\textsuperscript{-1}~Pa and 10\textsuperscript{6}~Pa. While this is not the most flexible model available (due to the development of two stream approximations \citep{HengTwoStreamRadTransfer,HELIOS_TwoStreamRadTransfer} or parametric models such as \cite{GuillotTempProfile}), it suits the purpose of this investigation since we are interested in the accuracy of a retrieval using common assumptions. Additionally, we use a temperature slightly cooler than the calculated equilibrium temperature in the literature (e.g. in \cite{WASP39b_EqTemp}) as isothermal profiles have been shown to retrieve lower temperatures than equilibrium temperatures by several studies including those by \cite{EffTempErrors} and \cite{Eff_Eqm_TempRetrievals}. 

As with the isothermal temperature model, chemical abundances are assumed to be constant throughout the atmosphere. More accurate profiles such as those using equilibrium chemistry (discussed in \cite{Al-Refaie2022_TauRexChemComparison}) or the two-layer model presented in \cite{Changeat_TwoLayerChem} are now available and the issue of day-night heterogeneity has been discussed in \cite{ChemistryBias}. Consideration of the effect of more complex chemical models is left to future investigations.

The full set of inputs used in the simulations can be found in table~\ref{table:Sim}. All models were generated using the forward modelling capabilities of TauREx\,\citep{TauREx1,TauREx2,TauREx3}. Examples of the three simulations (at varying resolutions) can be found in figure~\ref{fig:Simulations}.

For each model, simulations at varying resolution and with varying photometric errors were produced. The binning procedure used was to generate a spectrum of constant resolution across the wavelength coverage of JWST's NIRSpec PRISM using the binning functionality already present in the TauREx code. Errors were then simply set to be constant for each point across the grid. We give consideration to the limitations in these simulations and their simple error and resolution profiles in section~\ref{sec:BinFactorRes} where we employ an alternative binning method.

The resolution grid used steps of 10 from R~=~10 to R~=~100 and steps of 100 between R~=~100 and R~=~1000. The errors on the spectra followed steps of 25\,ppm between a minimum of 50\,ppm and a maximum of 200\,ppm. These values were chosen since the JWST ERS spectrum presented in \cite{NIRSpecPRISM_Wasp39b} has an average resolution of approximately 137 and an average error of approximately 111\,ppm across its data. Thus, the chosen grid includes and extrapolates beyond this point.

Given the large number of spectra this grid yields (and the additional factor of three enforced by the study of three different instances of cloud decks) we elect to perform retrievals on models without additional scatter applied to simulate observational noise. Following the central limit theorem, \cite{Feng_NoNeedToScatter} show that, when noise instances are drawn from a Gaussian distribution the two available methods, simulating thousands of randomised noise instances or taking a noise-free instance of the simulated spectrum, are equivalent in the results they yield except that the noise-free method may produce predictions which are better centred on the true value than they might otherwise be; the shapes of the posterior distributions remain unchanged.  

\subsection{Retrieval set up}
\label{sec:RetrievalSetUp}

For each of the retrievals conducted in the initial investigation, we used the Nested Sampling\,\citep{Skilling_NestedSamp} code, Multinest, through its PyMultinest implementation\,\citep{MultiNest_1,MultiNest_2,MultiNest_3,Buchner2014_PyMultinest} and 500 live points were set. In each case there were seven parameters to fit: planetary radius, effective temperature, molecular abundances of H\textsubscript{2}O, CO, CO\textsubscript{2} and SO\textsubscript{2} and the cloud deck pressure. Note that Na is not fit for in these analyses. Table~\ref{table:Prior} provides all of the corresponding prior ranges used in the investigation.

\begin{table}
    \centering
    \begin{tabular}{c|c|c}
    \toprule\toprule
        \textbf{Parameter} & \textbf{Prior range} & \textbf{Prior type}\\
        \midrule
        Planetary Radius (R\textsubscript{Jup}) & [1,1.5] & Uniform\\
        Effective Temperature (K) & [500,1500] & Uniform\\
        Log\textsubscript{10}(Chemical abundances) & [-1,-8] & Log uniform\\
        Cloud deck (Pa) & [10\textsuperscript{-3},10\textsuperscript{5}] & Uniform\\
    \end{tabular}
    \caption{The input priors are all uniform or log uniform (for chemical abundances). Each contains the 'true' value of the parameter used in the simulation except in the 'No cloud' case.}
    \label{table:Prior}
\end{table}

\subsection{Additional investigations}
\label{sec:AdditionalInvest}

\begin{figure}
    \includegraphics[width=\columnwidth]{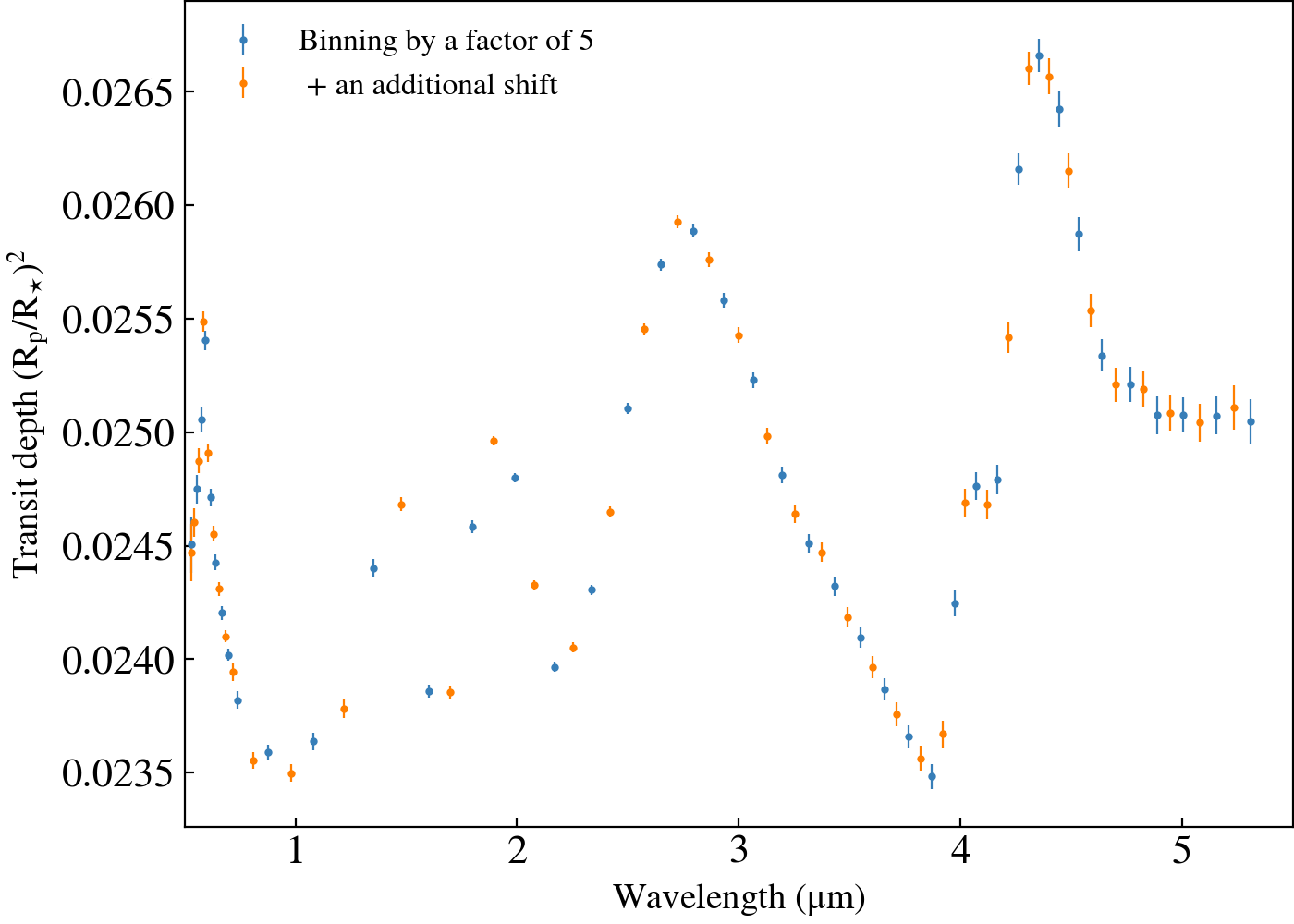}
    \caption{The different spectra obtained by binning the simulation of a planet like WASP-39b to the wavelength grid of the JWST NIRSpec PRISM data from the ERS program\,\citep{NIRSpecPRISM_Wasp39b} reduced by retaining every fifth point in the original wavelength grid with and without an additional shift.}
    \label{fig:ShiftedGrid}
\end{figure}

Having completed the basic retrievals (as described in the previous subsection), several additional tests were conducted. These tests aimed to study possible sources of bias based on the set up of our simulation and our retrieval strategies.

\begin{figure*}
    \includegraphics[width=\textwidth]{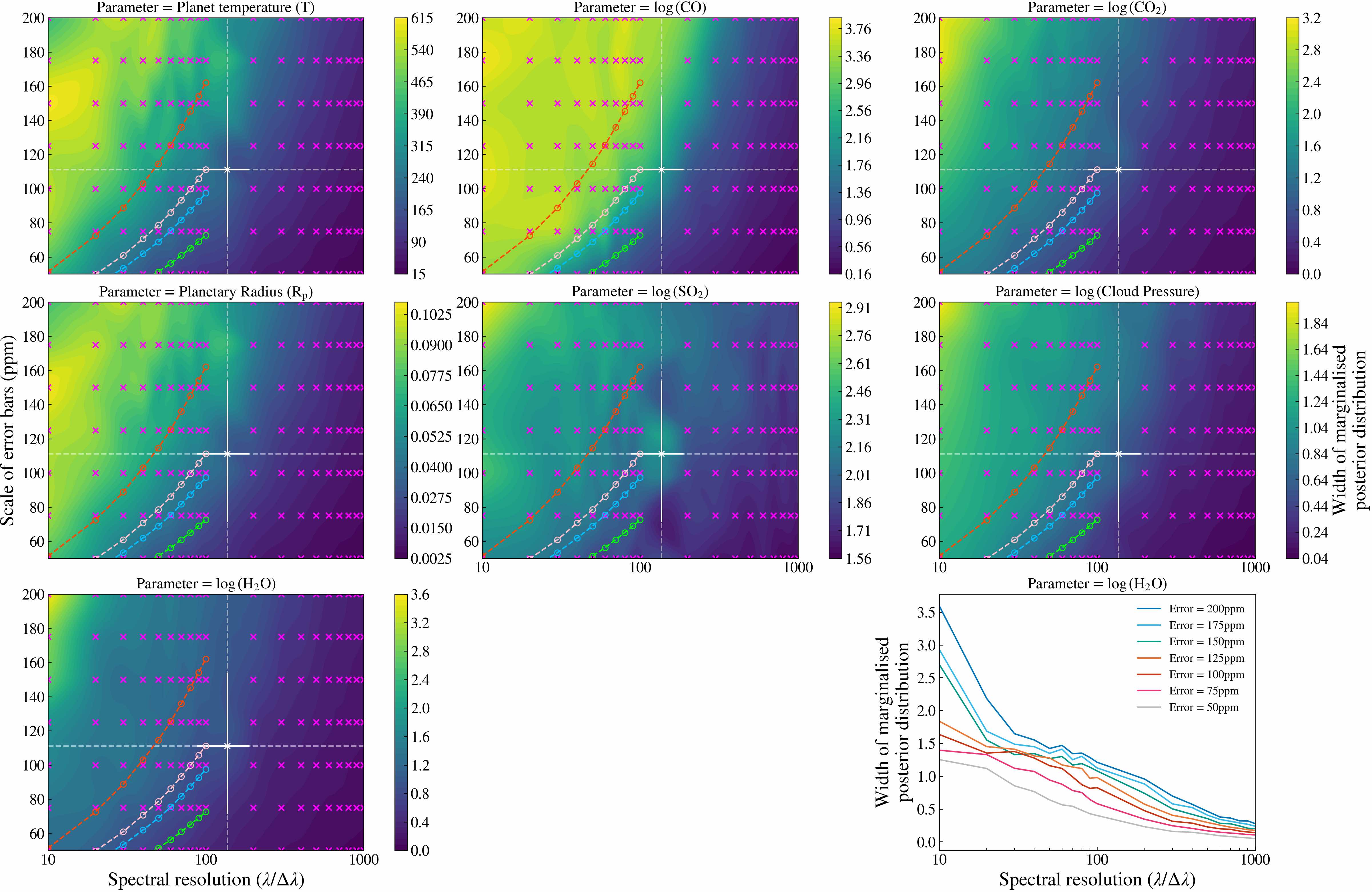}
    \caption{Posterior width maps for the case of a high cloud deck. In each subplot (labelled by which parameter is represented) the pink crosses mark the position of data from retrievals. Interpolation has been used to fill in the rest of the grid and generate a map. Retrievals are spaced by R\,=\,10 below R~=~100 and by 100 between R~=~100 and R~=~1000. The colour bar associated with each map describes the width of the marginalised posterior distribution obtained in retrievals for the given parameter. The white arrow marks the approximate position of the JWST ERS data in the map. The error bar on this point is set by the 25\textsuperscript{th} and 75\textsuperscript{th} percentiles of the variation in resolution and error in the input spectrum. The blue, green, red and pink dashed lines plot potential binning paths under the assumption that each bin (before binning) contains an equal number of photon counts. The start position is set by the average resolution of the ERS data and the average error in the shortest (blue), middle (green) and longest (red) wavelength regions in the data. The pink dashed line plots the potential binning path from the approximate position of the ERS data point, taking the average error over the whole spectrum. The lower right plot in the grid shows the same data as the lower left but as a series of lines.}
    \label{fig:HighCloudWidthMap}
\end{figure*}

\textbf{Bias due to binning method:} To investigate the impact of our choice of binning strategy, two additional types of binning grid were introduced. The first additional type of grid was based on the FIREFLy\,\citep{FIREFLy} reduction of the JWST data\,\citep{NIRSpecPRISM_Wasp39b} and in these cases the resolution is not constant across the spectrum. However, we are limited in the maximum resolution we can achieve by the resolution of the input spectrum as we can only bin down from the starting point of the input data. We scale down the resolution of these grids by retaining only every N\textsuperscript{th} point in the spectrum (where N is an integer between 2 and 8). In these cases, we use the mean resolution and photometric error across the spectrum to plot them in subsequent analyses. These tests were designed to validate our results from constant resolution grids by comparing them to more instrument specific examples. Example plots of these spectra can be seen in figure~\ref{fig:BinFactorSpec}.

For this first type of grid, we extend the analysis further by comparing predictions when binning directly from the resolution of the forward model to the desired wavelength grid and when first binning to the observation's (ERS data's) resolution as an intermediate step. 

The second type of grid was the same as the first but for each point in the wavelengths grids, a minor shift was added. This shift was positive in wavelength space and applied to the grid prior to binning the spectrum. The value of the shift was set such that the new points in the `shifted' grids fell between the points in the original, `unshifted' grids. This was achieved by adding half the difference in wavelength between subsequent points to the shorter wavelength point. An example of these spectra is shown in figure~\ref{fig:ShiftedGrid}.

\textbf{Bias due to model complexity:} Taking a few cases to sparsely sample the resolution-error grid, retrievals were re-run while fixing all of the chemical abundances other than that of H\textsubscript{2}O to their true values. We then fit only for planet radius, effective temperature, the abundance of H\textsubscript{2}O and the cloud deck pressure in the retrieval to perform a preliminary investigation into the dependence of the results on model complexity. In doing so, we wish to investigate whether or not the bias in predictions is reduced by knowing the log-abundances of other molecules a-priori and to what extent this depends on the resolved lines in the water absorption bands.

\textbf{Bias due to stochastic sampling:} Finally, we check for consistency in the results by running repeat instances of a subset of the retrievals. The aim of this study was to verify the stability of the sensitivity maps that we produce considering the fact that nested sampling is a stochastic process.

\begin{figure*}
    \includegraphics[width=\textwidth]{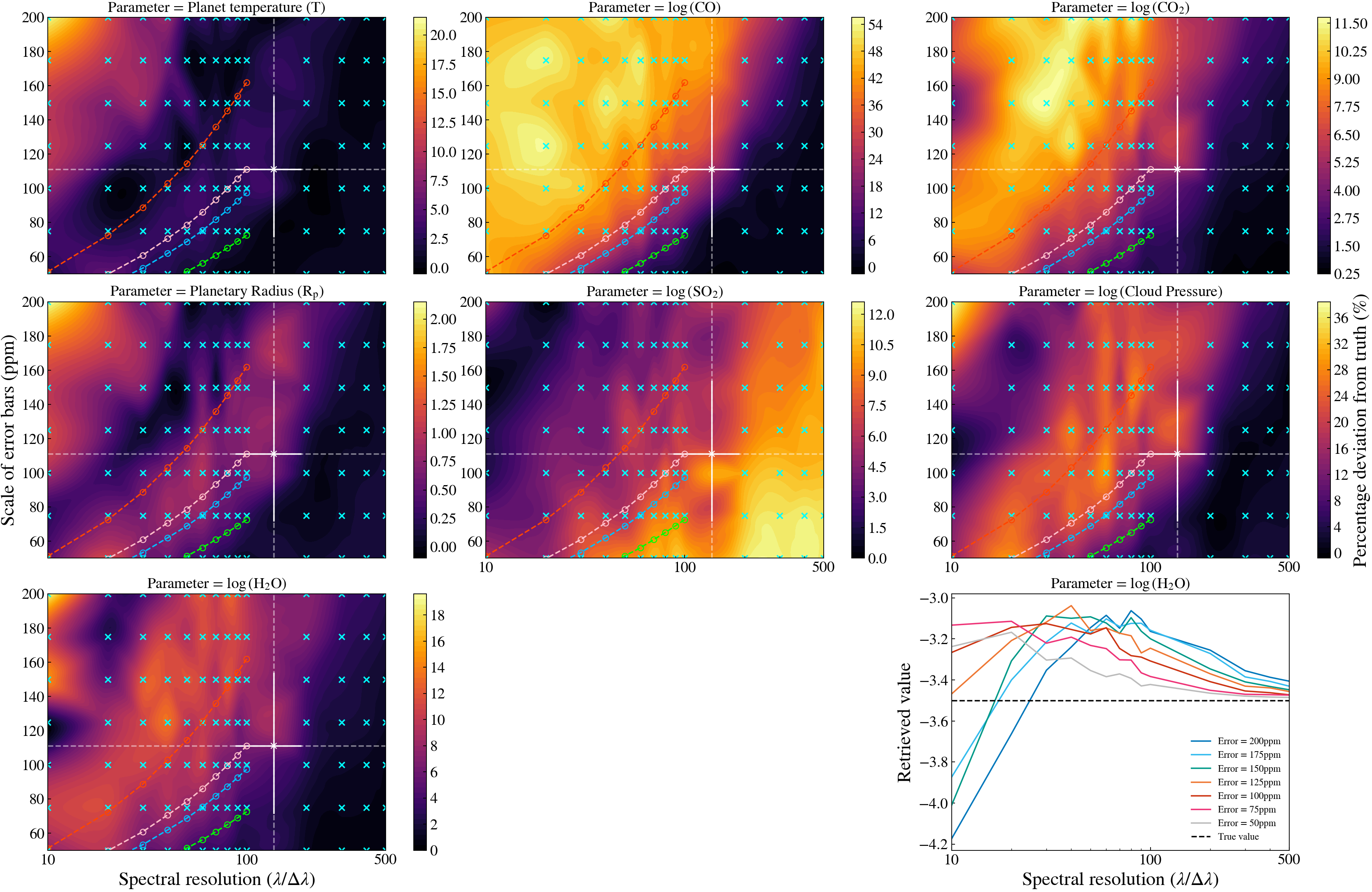}
    \caption{Same as figure\,\ref{fig:HighCloudWidthMap} but for the deviation of the mean value from the ground truth of the retrieved parameters.
    The colour bar associated with each map describes the percentage deviation of the value obtained in retrievals from the true value of that parameter (input to the simulation). The bottom panel shows a collapsed version of the grid for $\log$(H\textsubscript{2}O) across the entire resolution grid. Note that the y-axis uses the absolute value of the retrieved abundance rather than the deviation. The input value is marked by the black dashed line.}
    \label{fig:HighCloudSenMap}
\end{figure*}

\section{Results: Constant resolution-error grid}
\label{sec:Results}

Without any quantitative study, we would expect that as the resolution of the spectrum increased and the scale of the photometric error decreased, the retrieval would more confidently and more accurately converge on the `true' value for each parameter. Increasing resolution provides better coverage across the instrument's wavelength range and decreasing the size of errors offers us greater confidence in the levels of transmission in the spectrum. However, certain cases in this analysis showed evidence of additional structure in their sensitivity maps.

Below, we present %uncertainties
maps focusing on three separate but related aspects: 1) Parameter estimation uncertainty; 2) Induced biases in the mean retrieved values; and 3) Change in correlations in the conditional posterior distribution.

\subsection{Retrieval confidence}
\label{sec:PostWidth}

Before assessing the bias of parameter estimations in the retrievals, we consider the confidence with which the sampling returns predictions, i.e. the retrieved uncertainty in the free-parameters fitted. In general, we observe a logarithmic decrease in the width of the posterior distribution (measured by the extent of the 1\,$\sigma$ Bayesian credible interval on the marginalised posterior) as the resolution of the spectrum increases. A map for this metric for the high cloud case can be seen in figure\,\ref{fig:HighCloudWidthMap} (with equivalent plots for the low and no cloud cases demonstrating similar trends included in Appendix\,\ref{Appendix:AdditionalPlots}, figures\,\ref{fig:LowCloudWidthMap} and \ref{fig:NoCloudWidthMap} respectively). At the high resolution end, this means that there is very little reduction in the size of the confidence interval for resolutions above R\,=\,100. In all cases, we observe parameter uncertainties to be a smooth and continuous function of SNR and resolution, as expected.

\begin{figure*}
    \includegraphics[width=\textwidth]{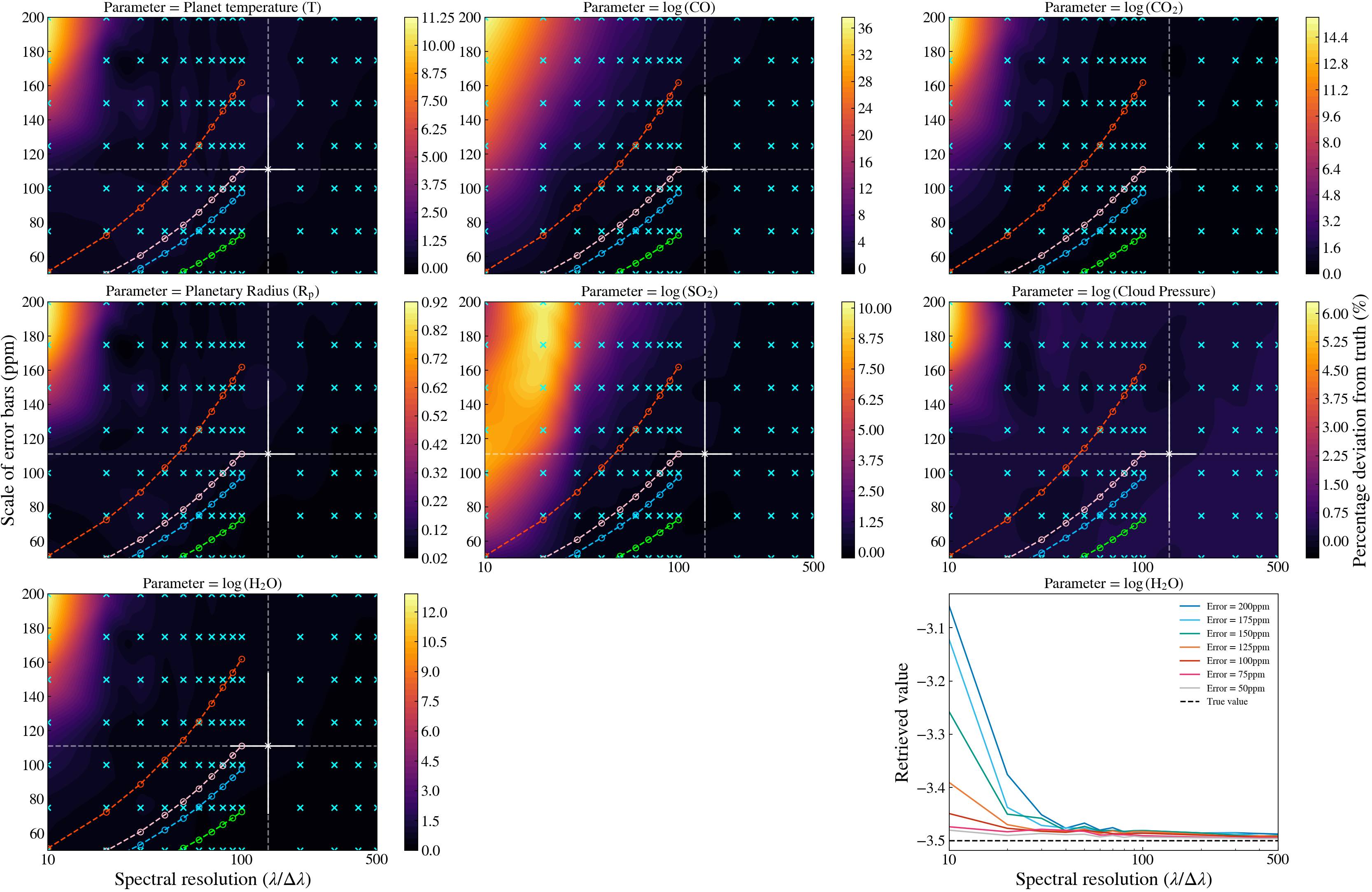}
    \caption{Sensitivity maps for the case of a low cloud deck. In each subplot (labelled by which parameter is represented) the blue crosses mark the position of data from retrievals. Interpolation has been used to fill in the rest of the grid and generate a map. A log scale is used for the spectral resolution (x axis) since these data are spaced by 10 below R~=~100 and by 100 between R~=~100 and R~=~500. The colour bar associated with each map describes the percentage deviation of the value obtained in retrievals from the true value of that parameter (input to the simulation). The white arrow marks the approximate position of the JWST ERS data in the map (including error bars set by the 25\textsuperscript{th} and 75\textsuperscript{th} percentiles of the variation across the observed spectrum in resolution and error bar size). The green dashed lines plot potential binning paths under the assumption that each bin (before binning) contains an equal number of counts. The pink dashed line plots the potential binning path from the approximate position of the ERS data point under the same assumptions. The bottom panel shows a collapsed version of the grid for $\log$(H\textsubscript{2}O) across the entire resolution grid. Note also that the y-axis uses the absolute value of the retrieved abundance rather than the deviation. The input value is marked by the black dashed line.}
    \label{fig:LowCloudSenMap}
\end{figure*}

Considering only this information, it would suggest there is very little to be gained from retrieving at higher resolutions for the broad band absorbers considered here. One would not expect to achieve any significant improvement in the confidence of one's predictions at the expense of additional computational cost. Thus, analysis of the posterior width maps alone would suggest that it is beneficial to bin down higher resolution spectra to R\,=\,100 as a limiting case.

Furthermore, the coloured lines in the lower right panel of figure~\ref{fig:HighCloudWidthMap} show the trends of the width of the marginal posterior distributions for different photometric errors on the spectra as resolution increases. We observe convergence of these lines suggesting that the confidence becomes insensitive (within reasonably expected SNRs) to the size of error bars on the input spectra at high resolutions.

Here, the data enter into a resolution limited region and this suggests that, at higher resolutions, if we assume that the scale of the photometric error on a spectrum can be decreased with longer observing time (i.e. multiple observations of the target), less time would be required for observations since there is little improvement to be gained by substantially reducing the size of error bars. In this regime, the resolution rather than the SNR has the greater influence on the precision of our predictions. However, we should still give consideration to both resolution and SNR when planning observations as significant uncertainty on our data could still lead to uninformative predictions. Furthermore, we treat the posterior widths as only the theoretical lower bound on the widths that would be retrieved for real data since we use the same forward model in producing our simulations and performing the retrieval.

\subsection{Bias maps (10\,$\le$\,R\,$\le$\,500)}
\label{sec:SensitivityMaps}

\begin{figure*}
    \includegraphics[width=\textwidth]{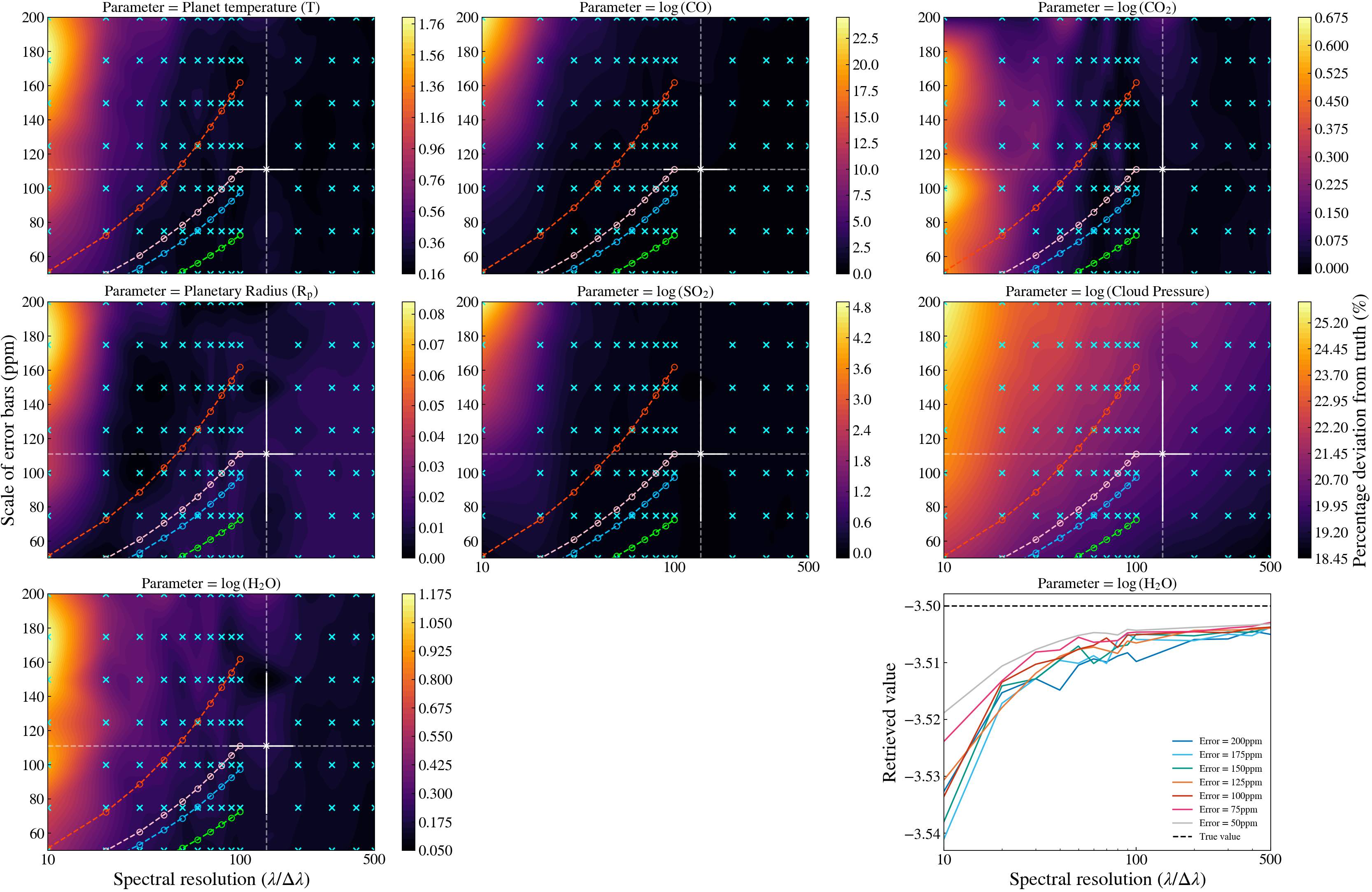}
    \caption{Sensitivity maps for the case of a no clouds included in the simulation. In each subplot (labelled by which parameter is represented) the blue crosses mark the position of data from retrievals. Interpolation has been used to fill in the rest of the grid and generate a map. A log scale is used for the spectral resolution (x axis) since these data are spaced by 10 below R~=~100 and by 100 between R~=~100 and R~=~500. The colour bar associated with each map describes the percentage deviation of the value obtained in retrievals from the true value of that parameter (input to the simulation). The white arrow marks the approximate position of the JWST ERS data in the map (including error bars set by the 25\textsuperscript{th} and 75\textsuperscript{th} percentiles of the variation across the observed spectrum in resolution and error bar size). The green dashed lines plot potential binning paths under the assumption that each bin (before binning) contains an equal number of counts. The pink dashed line plots the potential binning path from the approximate position of the ERS data point under the same assumptions. The bottom panel shows a collapsed version of the grid for $\log$(H\textsubscript{2}O) across the entire resolution grid. Note also that the y-axis uses the absolute value of the retrieved abundance rather than the deviation. The input value is marked by the black dashed line.}
    \label{fig:NoCloudSenMap}
\end{figure*}

\begin{figure*}
    \includegraphics[width=\textwidth]{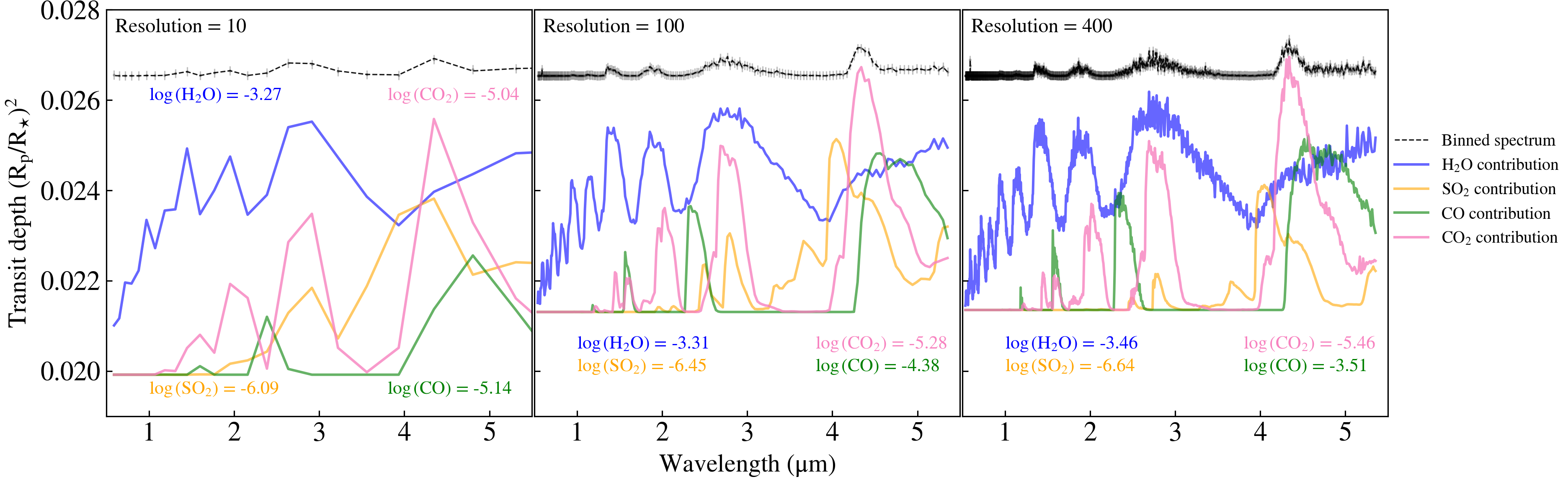}
    \caption{Plots of the retrieved spectra in the case of a high cloud deck with errors of 100\,ppm. Below a resolution of R\,=\,500, only SO$_{2}$ shows a reduction in the accuracy of it's predictions. Shown here are the contributions to the retrieved spectra at resolutions of R\,=\,10, R\,=\,100 and R\,=\,400. Coloured lines indicate retrieved contributions from individual molecules while the dashed black line presents the combined spectrum binned to the resolution of the input data.}%Note that only contributions from H$_{2}$O, SO$_{2}$ and CO are shown for clarity.}
    \label{fig:SpecContributions}
\end{figure*}

Next, we turn our attention to the accuracy of our grid of retrievals. Figures\,\ref{fig:HighCloudSenMap},~\ref{fig:LowCloudSenMap} and~\ref{fig:NoCloudSenMap} show the sensitivity maps for each parameter in retrievals for the high, low and no cloud scenarios respectively. In these plots, the colour bar corresponds to the percentage deviation of the retrieved value from the true value (as defined in section~\ref{sec:Methodology}).

On each of the maps, blue crosses mark the positions of the retrievals in resolution-error space. The horizontal and vertical dashed lines mark the position of the average error and the average resolution of the ERS observation. The solid lines at their intersection demonstrate the extent of the variation of data error and resolution element across the ERS spectrum by marking the 25\textsuperscript{th} and 75\textsuperscript{th} percentiles for each value. Additionally, `binning paths' (as defined in section~\ref{sec:BinMotivation}) have been plotted in pink starting from the ERS data point and in blue, green and red starting from the average error on the spectrum at the shortest, middle and longest wavelength bins marked by coloured arrows on the left panel of figure\,\ref{fig:Simulations}.

Note that these maps only plot the data up to a resolution of 500 since there is less variation beyond this point and the inclusion of high resolution retrievals can mask the structure seen at lower resolutions if included in the maps. The high resolution end of the grids can be seen in appendix~\ref{App:HighResMaps}, figures~\ref{fig:HighCloudSenMap_HighRes}, \ref{fig:LowCloudSenMap_HighRes} and \ref{fig:NoCloudSenMap_HighRes} for the high, low and no cloud cases respectively.

In the case of high clouds (shown in figure\,\ref{fig:HighCloudSenMap}) there is a significant amount of structure present in the bias maps across all resolutions and error scales. Notably, the maps of log-abundances of CO or CO\textsubscript{2} and that of SO\textsubscript{2} appear to present inverse patterns to one another. 

Shown in figure~\ref{fig:SpecContributions}, the most significant cause of this inverted structure appears to be the underestimation of the CO\textsubscript{2} abundance. The retrieved values of CO\textsubscript{2} vary by 0.42 in log-abundance between R\,=\,10 and R\,=\,400. In part, this is due to the overlap in the spectra of CO\textsubscript{2} and SO\textsubscript{2} at low resolutions. At R\,=\,10, both exhibit peaks at 4.35\,\textmu m and 2.91\,\textmu m. However, as resolution increases, more of the structure in and around these features is sampled. Of particular importance is the increase in sampling between the bins of 3.93\,\textmu m and 4.35\,\textmu m for the SO\textsubscript{2} spectrum. This reveals that the peak actually occurs at a slightly shorter wavelength for SO\textsubscript{2} than it does for CO\textsubscript{2} (4.05\,\textmu m rather than 4.35\,\textmu m).

SO$_{2}$ is the only parameter of those retrieved where we observe a different trend and the bias seems to worsen with increasing resolution in figure\,\ref{fig:HighCloudSenMap}. This is an artifact of the low volume mixing ratio of SO$_2$ in this simulation and the fact that SO$_2$ and CO$_{2}$ emission features are not separable at low resolution and that SO$_{2}$ can only be retrieved as an upper limit in our simulation. 
If we consider the width of the SO$_{2}$ marginal posterior distribution (see figure~\ref{fig:HighCloudWidthMap}), we can see that the values of the log-abundance of SO$_{2}$ are less well constrained and we are only ever able to place an upper limit on its prediction whereas the log-abundance of other molecules become well resolved at higher resolutions. (This is also shown in the corner plot included in appendix~\ref{App:CornerPlots}.)

Maps for all of the other parameters show their most significant differences when the resolution is reduced below R~=~100 and this holds true for the other cloud scenarios as well (see figures \ref{fig:LowCloudSenMap} and \ref{fig:NoCloudSenMap}). Though their patterns are generally matching, it is worth noting the difference in the scales of these bias maps. In general, planetary radius is well constrained and only deviates by around 2\,\% in the worst case scenario. However, the molecular abundances can experience deviations up to about 30\,\% across the different molecules and cloud scenarios (with $\log$(CO) sitting as an outlier with a deviation as high as 55\,\% in the high cloud case).

The common structure observed does follow the expectation that spectra with low resolution and at high photometric error (i.e. the top left corner of each panel) would be the worst retrieved.

The bottom right panel of each map presents a collapsed version of the maps for the log-abundance of H\textsubscript{2}O where each line shows the retrieved value at various spectral noise levels. This allows us to see if the retrieval is inaccurate due to over- or under-estimations in the parameters.

We find that, in the case of H\textsubscript{2}O, the log-abundance is more often over-estimated in cases including clouds but under-estimated when clouds are absent. However, it is worth noting the change in scale between the cases including clouds and the no cloud case. With clouds, the prediction for the log-abundance of H\textsubscript{2}O varies by approximately $\pm$~0.4 but in the case without clouds the magnitude of the deviation is about ten times less ($\pm$~0.04). 
%\textbf{This is likely due to the fact that the retrieval still fits a cloud deck in the no cloud scenario because the priors are left the same for each retrieval. Thus, it is forced to fit a cloud layer at the lowest possible altitude (as allowed by its priors) and compensates for this by reducing the abundance of H\textsubscript{2}O.}

\subsection{Correlation analysis}
\label{sec:Correlation}

\begin{figure}
    \includegraphics[width=\columnwidth]{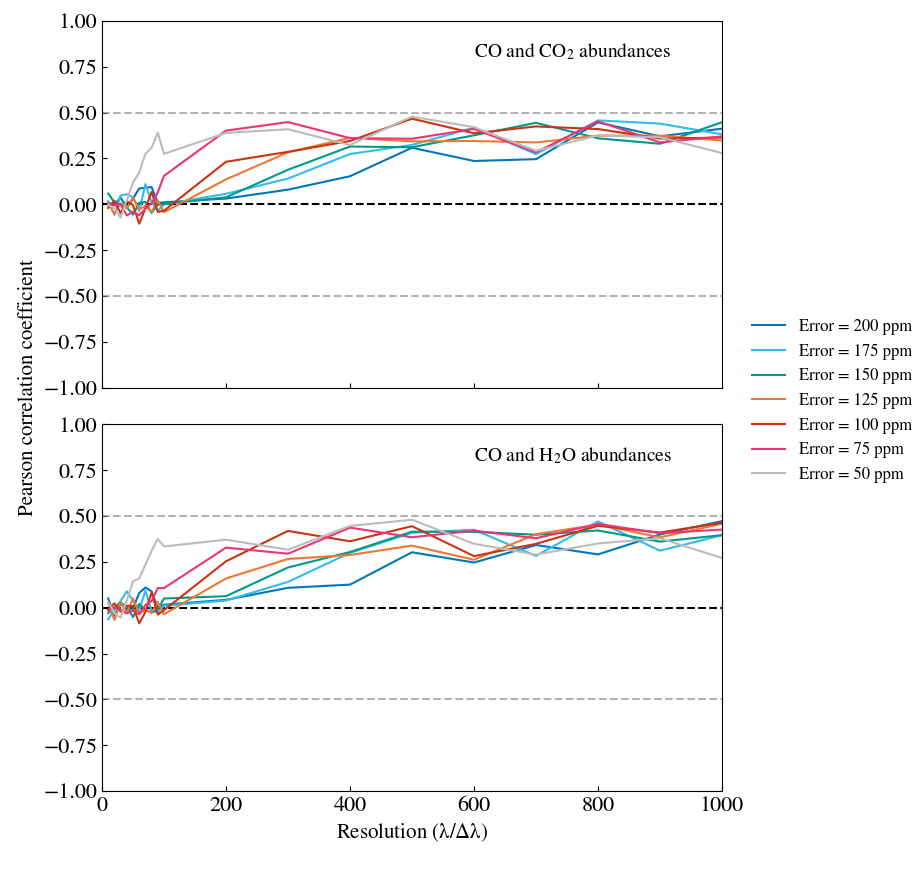}
    \caption{The correlation between specific retrieved parameters in the case of a spectrum with a high cloud deck. The parameters under investigation are marked by the text in each panel and the different coloured lines show the dependence of these trends on the scale of photometric error under investigation. Note that only the pairings of parameters where there is a significant change in the correlation across the resolution grid are displayed here.}
    \label{fig:HighCloudCor}
\end{figure}

\begin{figure}
    \includegraphics[width=\columnwidth]{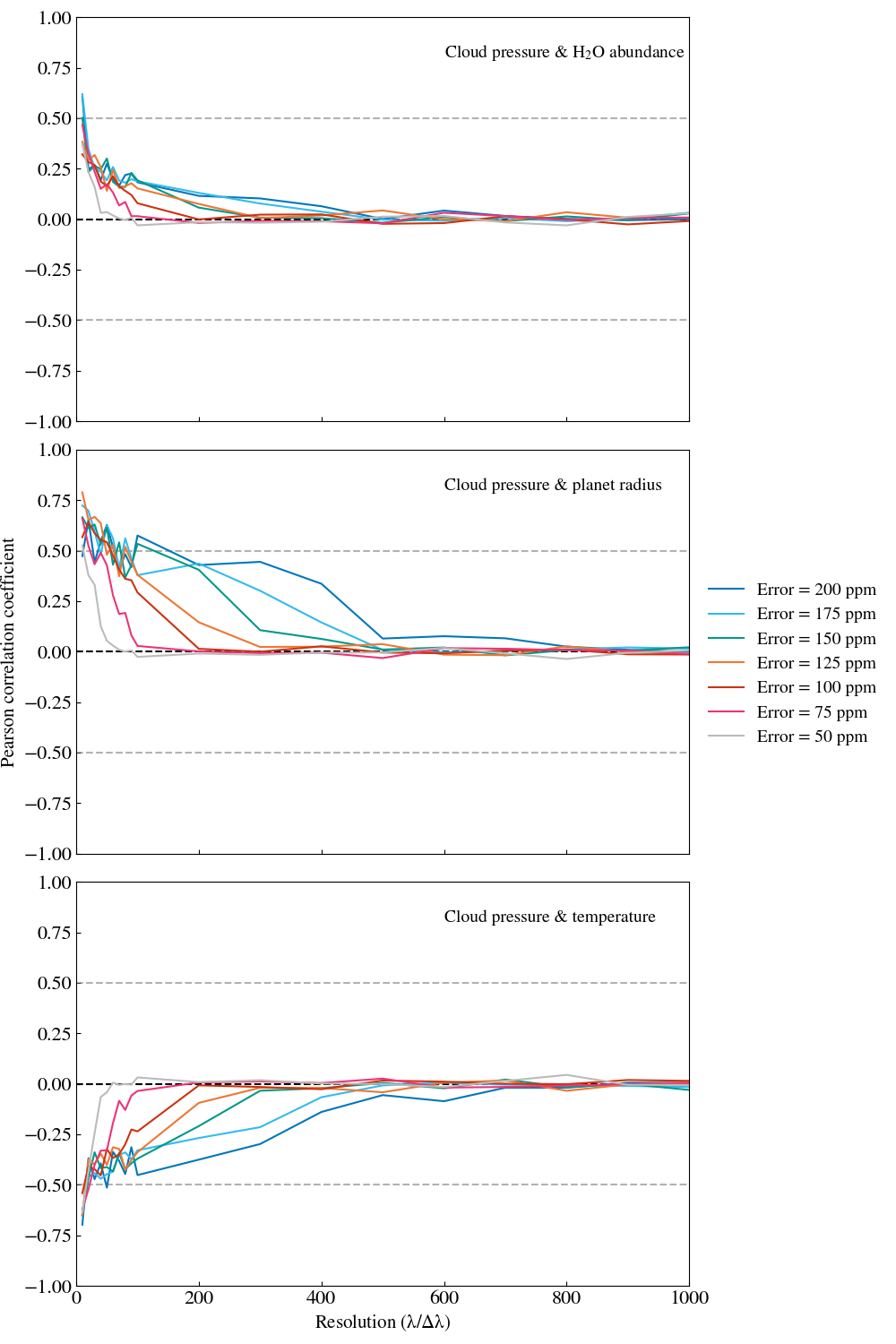}
    \caption{Same as in figure~\ref{fig:HighCloudCor} but now focusing on pairings of retrieved parameters in the case of spectra with a low cloud deck. In this case, the pressure of the cloud deck is the parameter that causes the most significant changes in the correlations.}
    \label{fig:LowCloudCor}
\end{figure}

For each possible pairing of parameters, we calculate the Pearson correlation coefficient on their sampling traces. In doing so, we are studying the fitted parameter degeneracies found in the conditional posterior distributions. As individual parameters become more or less retrievable as a function of SNR and resolution change, we expect the underlying degeneracies in the radiative transfer solutions to reflect this. Different trends are seen for each cloud case. Of particular interest are cases where the correlation changes as a function of resolution. Several examples are discussed in this section but we give consideration to the high cloud case first in order to further investigate the structure seen in the bias maps (figure~\ref{fig:HighCloudSenMap}).

For the case of high clouds the most significant changes are found with the fitting of the log-abundance of CO and its dependence on the log-abundances of CO$_{2}$ or H$_{2}$O. Plots of these trends are given in figure~\ref{fig:HighCloudCor}'s top and bottom panels respectively. In both cases, the correlation between the traces increases with increasing resolution. At our lowest resolution in the retrieval grid, we find that there is little to no resolvable correlation between these parameters. However, at a resolution of 1000, the value approaches +0.50 as we are better able to resolve the features in the spectrum and when the value of correlation stabilises, we find the natural degeneracies in the retrieval (rather than those caused by an undersampling of the spectrum).

It is also worth noting that the spectra with smaller uncertainties appear to see a more rapid increase in the correlation between parameters as spectral resolution increases. Thus, with smaller uncertainties, we are able to resolve the natural degeneracies between parameters at a lower spectral resolution since we retain more information in the spectrum at any given resolution step.

In the case of our low cloud deck, we observe a reduction in the correlation between several parameters with increasing spectral resolution. The three most prominent cases of this effect are shown in figure~\ref{fig:LowCloudCor}. This shows clearly the impact spectral resolution has in mitigating model degeneracies.

This is the opposite effect to that seen in the high cloud case where correlation increased. One similarity comparing to the trends in the high cloud case is the dependence on photometric error. We note in the low cloud cases that the smaller the photometric error, the more rapid the change in the correlation with increasing resolution for the reasons mentioned previously. 

With the low cloud spectra, the parameter which experiences the greatest change in correlation is the cloud top pressure. This is degenerate with the temperature, planetary radius and H$_{2}$O log-abundance at low resolution but parameters become disentangled at higher resolutions as is demonstrated by the reduction in the magnitude of the Pearson correlation coefficient.

An additional interesting feature of these correlation plots is the scale at which the values stabilise and we uncover the natural degeneracies of the model fit. The minimum resolution at which all of the lines in plots~\ref{fig:HighCloudCor} and \ref{fig:LowCloudCor} reach a stable value is approximately 500. However, if we compare this to the resolution at which the values of our input parameters are well predicted in the bias maps, we can see that, even in the least favourable conditions (the high cloud deck), parameters are well predicted above a resolution of 200. This means that we require a much higher resolution to reveal degeneracies between parameters than we do to fit the correct values in a retrieval. 

\section{Results: Binning from ERS data}
\label{sec:BinFactorRes}

In order to validate our methods, we also consider the case of retrieving from a binned version of the JWST ERS data as described in section~\ref{sec:AdditionalInvest}.
%for the NIRSpec PRISM instrument observing WASP-39b \citep{NIRSpecPRISM_Wasp39b}. These spectra were binned by taking the original wavelength grid of the FIREFLy reduction of the spectrum and reducing it by only retaining every N\textsuperscript{th} point (where N is an integer between 2 and 8 which we refer to as a binning factor). The original data were then binned to this new grid using TauREx which propagates the error bars from the original ERS data as input (averaging 111\,ppm across the spectrum). Examples of these spectra are shown in figure~\ref{fig:BinFactorSpec}.
\begin{table}
    \centering
    \begin{tabular}{c|c|c}
    \toprule\toprule
        \textbf{Binning factor} & \textbf{Average spectral resolution} & \textbf{Average error (ppm)}\\
        \midrule
        1 & 137 & 110\\
        2 & 68 & 69\\
        3 & 45 & 64\\
        4 & 34 & 53\\
        5 & 27 & 51\\
        6 & 22 & 46\\
        7 & 19 & 43\\
        8 & 17 & 40\\        
    \end{tabular}
    \caption{The corresponding mean spectral resolution and photometric error for the spectra when binned down from the JWST ERS observations by a given binning factor. Note that the resolution and error values have been rounded to the closest integer value.}
    \label{table:BinFacError}
\end{table}
The percentage deviations for each parameter in retrievals on these spectra are shown in figure~\ref{fig:BinFactorDev}. Note that there is no separation by error scale in this case since the errors are set by the propagation of error bars from the ERS observations. Thus, they have pre-defined values for each wavelength grid. A table of the average error for each grid can be found in table~\ref{table:BinFacError}.

This method of binning perhaps better represents JWST NIRSpec data as it retains some of the variation across the spectra considering both error and resolution whereas our other simulations use constant resolution and photometric error on every point. However, this alternative technique also bears limitations. Most notably, you cannot bin up from the input data so there is a maximum set on the resolution that can be used in such analysis.

\subsection{Comparison to constant resolution-error grid results}

\begin{figure}
    \includegraphics[width=\columnwidth]{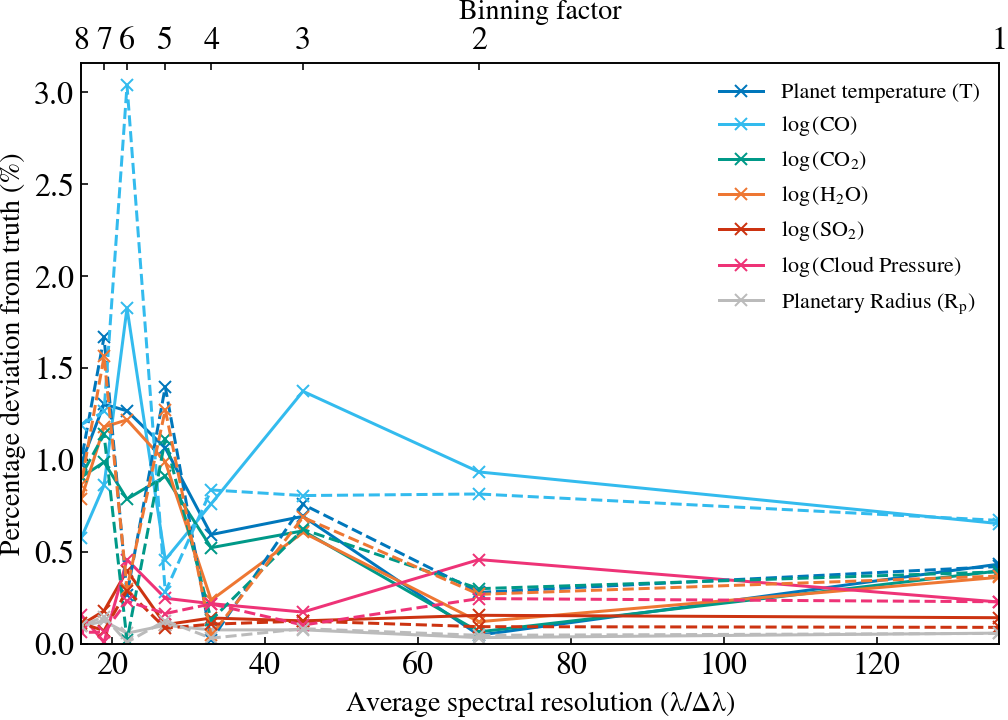}
    \caption{The percentage deviation of retrieved parameters from spectra obtained by binning down simulations of the JWST ERS data for WASP-39b using the NIRSpec PRISM \citep{NIRSpecPRISM_Wasp39b} in the low cloud case. The wavelength grid was constructed by retaining only every N\textsuperscript{th} point in the spectrum where N is the `Binning factor'. Each line represents a different parameter and the top and bottom x-axes quantify the level of binning through the binning factor and the average resolution  respectively. All values for average resolution are obtained by considering the mean value across the spectrum. Solid lines show data from retrievals where the simulation was first binned to the wavelength grid of the ERS data and then binned further whereas the dashed lines bin directly from the native resolution of the TauREx simulation to the appropriate reduced grid.}
    \label{fig:BinFactorDev}
\end{figure}

In figure~\ref{fig:BinFactorDev} we observe a general trend of increasing accuracy with increasing spectral resolution as is expected. In most cases, there are some higher binning factors (corresponding to a lower spectral resolution) where predictions appear to show reduced biases for some parameters whilst increased biases for other. We attribute this to lucky sampling that is better suited to one parameter compared to others. 

To further investigate the influence of binning strategy on the predictions of retrievals, we investigate the consequences of multiple binning instances on a single dataset and compare to a single instance of binning. In figure~\ref{fig:BinFactorDev}, the dashed lines provide the predictions when data are binned to the desired wavelength grid directly from the native resolution of the TauREx forward model. Alongside these, the solid lines provide the results when binning from the native resolution to the ERS wavelength grid and then binning further to the desired grid. (Note that this means that the spectra at a binning factor of 1 are identical with both methods and, as expected, agree in their predictions.)

This additional test is designed to simulate the process of observation through the initial binning step (to the ERS wavelength grid) by simulating the noise induced by binning the spectra. Results between the two methods are largely consistent with some variation between the predictions stemming from their prediction uncertainties. In general we find that the scale of deviations agree between our retrievals in this test and our initial resolution-error grids and they follow the same broad trends. Additionally, our results indicate that the decrease in resolution is more significant in the accuracy of predictions than the corresponding decrease in photometric error since predictions worsen at lower resolutions (higher binning factors).

\subsection{Shifted wavelength grids}
\label{sec:ShiftedGrids}

\begin{figure}
    \includegraphics[width=\columnwidth]{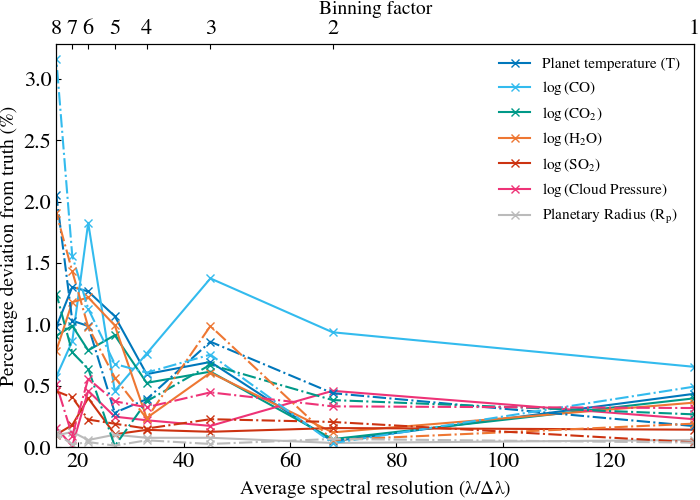}
    \caption{The solid lines represent the same data as in figure~\ref{fig:BinFactorDev}. However, the dot-dashed lines present results from retrievals where there initial wavelength grids of spectra used to retrieve the solid lines have had an additional wavelength shift applied prior to binning the spectrum. This shift is set point-by-point across the spectrum such that the resulting points in the new grid are halfway between the original point and the next in the un-shifted grid. See figure~\ref{fig:ShiftedGrid} for examples of these spectra.}
    \label{fig:BinFactorDev_withShift}
\end{figure}

As described in section\,\ref{sec:AdditionalInvest}, we have performed additional retrievals testing for the impact of a slight wavelength grid shift (see figure~\ref{fig:ShiftedGrid} for an example spectrum). We continue to see broad consistency with the results from `shifted' and `unshifted' grids. These results are shown in figure~\ref{fig:BinFactorDev_withShift} for spectra of the low cloud case (with similar plots for the high and no cloud spectra shown in appendix~\ref{App:BinFactor_shift}, figures~\ref{fig:HighCloudBinFac} and \ref{fig:NoCloudBinFac}).

There appear to be offsets between some parameters when retrieving from shifted and unshifted spectra but the same trend of worsening predictions with decreasing spectral resolution is observed at similar magnitudes. Hence trends described earlier are irrespective of the type of grid chosen. 

\section{Results: Additional tests}

The final tests in this investigation were used to validate our retrieval methodology and to investigate any bias emerging from model complexity or from the stochasticity of our sampling procedure. 

\subsection{Simplified retrievals and models}
\label{sec:ModelComplexityStudy}

In order to investigate the dependence of our results on model complexity we perform two additional types of retrieval. In the first case, we leave the base simulation as defined in section~\ref{sec:SimulationSetUp}. However, we change the fitting parameters in the retrieval. The log-abundances of all molecules other than H\textsubscript{2}O are fixed to their correct input values and retrievals are run while fitting only for temperature, planetary radius, cloud deck pressure and log-abundance of H\textsubscript{2}O.
\begin{table}
    \centering
    \begin{tabular}{c|c|c|c}
    \toprule\toprule
        \textbf{Type of} & \textbf{Molecule} & \textbf{Included in the} & \textbf{Fit in}\\
        \textbf{retrieval} &  & \textbf{input spectrum} & \textbf{retrieval} \\
        \midrule
        Simplified Retrieval & H\textsubscript{2}O & True & Free\\
                             & CO\textsubscript{2} & True & Fixed\\
                             & CO                  & True & Fixed\\
                             & SO\textsubscript{2} & True & Fixed\\
                             & Na                  & True & Fixed\\
        \midrule
        Simplified Model     & H\textsubscript{2}O & True & Free\\
                             & CO\textsubscript{2} & False & - \\
                             & CO                  & False & - \\
                             & SO\textsubscript{2} & False & - \\
                             & Na                  & False & -\\
        
    \end{tabular}
    \caption{The chemical input and fittable parameters for the simplified retrieval and simplified model tests. In the final column, `Free' indicates that the value is fit during the retrieval, `Fixed' Indicates that it is set to its correct value a-priori and `-' indicates that it is left out of the retrieval all together.}
    \label{table:SimplifiedRetrievals}
\end{table}

We call this a simplified retrieval. Here, we allow the retrieval to vary only a subset of the parameters that are present in the simulation. Thus, it is restricted in the possible combinations of parameters it can sample due to the reduction in the dimensionality of the prior space.

The second type we refer to as a simplified model. The simplified model refers to a case in which the base simulation only contains H\textsubscript{2}O as a molecular contribution to the atmospheric absorption. All the other chemical contributions (CO, CO\textsubscript{2}, SO\textsubscript{2} and Na) are neglected in these spectra. The retrieval is then run on this spectrum fitting for temperature, planetary radius, cloud deck pressure and log-abundance of H\textsubscript{2}O as in the previous test.

A breakdown of the chemical inputs and fitting parameters for these retrievals can be seen in table~\ref{table:SimplifiedRetrievals}.

In our results we find that there is little dependence on the model complexity under these tests. Figures of these data are shown in the appendices (appendix~\ref{App:Complexity}, figure~\ref{fig:OnlyH2OFit}). At low resolution, the original grid performs slightly worse than the simplified retrievals but towards higher resolutions, all retrievals converge on the true input parameters. Intuitively, this makes sense since the simplified retrieval has fewer parameters to deal with and, as such, converges on the true solution faster and with greater confidence. By fixing other molecular abundances to their correct values, we are removing the uncertainty that arises from degeneracies between the log-abundance of H\textsubscript{2}O and other molecular contributions.

\subsection{Repeat retrieval instances}
\label{sec:Repeat}

We run two tests on repeat retrieval instances. In the first case, we are testing the stability of our parameter predictions given the stochastic nature of the nested sampling procedure used. For this investigation we simply re-run retrievals from the original retrieval grid and compare the percentage deviations from each instance of the retrieval procedure.

Overall we find minimal difference when the retrieval is re-run. As expected, the difference between the two instances of retrievals reduces with increasing resolution since the confidence in predictions increases in this regime. See appendix~\ref{App:Repeat} for figures displaying the results of this additional investigation.

Then, we also test the variation in our results when we introduce some artificial scatter into the spectrum. We do this for spectra in the low cloud case across all resolutions. In the scattered spectra each data point is resampled from a Gaussian distribution centred on the original point and with a width set by the size of the 1\,$\sigma$ photometric error. We run three tests in this investigation using the best (smallest, 50\,ppm), worst (largest, 200\,ppm) and a mid-level (100\,ppm) photometric error.

Overall we find that, as with the previous test, the difference in the percentage deviation for each retrieval decreases with increasing resolution since the data are then more heavily influenced by their wavelength coverage than by their errors. There are larger discrepancies in these cases than in the repeat cases with no scatter, particularly at the low resolution end where we see differences of up to approximately 25\,\% in the worst case scenario (where the photometric error is 200\,ppm on every spectral point). This is due to the reliance of well constrained transit depths in the lower resolution spectra which is accounted for with a more finely sampled wavelength grid at higher resolutions. 

These data are available in the appendices (see appendix~\ref{App:Perturb}).

\section{Discussion and conclusions}
\label{sec:DiscAndConc}

The motivation for binning is indisputable. It is a useful tool when it comes to analysis of exoplanet atmospheres using Bayesian retrieval methods, reducing the necessary computing resources and improving the SNR of individual points in the spectra. However, binning below R~=~100, we observe a logarithmic increase in the width of the posterior distribution. Thus, even in cases where the correct abundance is retrieved, it's credible interval remains large and poorly constrained. It is necessary in such cases to consider how informative these results are as, for molecular abundances, these uncertainties can cover orders of magnitude. 

This is particularly true for high cloud cases where spectral features are muted. In these cases, reducing resolution of spectra comes at a significant cost and the region where atmospheric parameters can be well retrieved is shifted towards a higher resolution in our grid of resolution-error space. 

It is pleasing to see that, in the low cloud case which is the one which most accurately mimics the shape of the NIRSpec PRISM data observed in \cite{NIRSpecPRISM_Wasp39b} for WASP-39b, at the native resolution of the observation, the data fall into a relatively safe zone in the retrieval grid, suggesting the right conditions for near optimal retrievals. Our analysis suggests similar accuracy could be achieved with additional binning of this data set but the results will only be as good, not better, and, again, the uncertainty on the retrieved value would grow significantly.

While this initial investigation targets a specific case, the method should be applicable to other planets. We would expect planets of similar type (e.g. size and mass), orbital configuration (e.g. orbital period and separation) and with similar host stars to follow similar trends as those observed in this study. However, if such a planet were to have a high cloud deck, caution would be strongly advised with JWST's NIRSpec PRISM as the native resolution of the instrument sits right on the boundary of the well and poorly retrieved cases in resolution-error space. In this case it would be advisable to obtain additional transit observations to reduce the average photometric error in the initial data. This comment may also apply to studies of smaller planets where atmospheric features are muted by a reduced atmospheric scale height rather than a cloud deck. 

\subsection{Limitations and Future Work}
\label{sec:Limitations}

The main limitation of this study is the fact that it targets a specific case, namely, that of a planet similar to WASP-39b. All of the results presented in this study apply to the simulated planet described in the methods sections. How easily these results can be generalised across a wider exoplanet population remains to be seen.

Another limitation emerges from the wavelength range and resolution-error grid considered in this study. For most of the investigations presented in this initial work, we use spectra of constant resolution and equal error bars across their entire wavelength range. This is simplistic but informative as a first analysis. However, we are pleased to see that our study of alternative binning strategies seem to present results which are largely in agreement with the data presented in our bias maps.
Results shown are specific to the JWST NIRSpec PRISM mode. Other instruments feature different spectral resolutions and wavelength coverages and results presented here may not be directly applicable to those modes.

We here presented a uniform binning grid across the full spectral range. Wavelength-dependent binning that is sensitive to the presence of spectral features may be a more optimal approach. Such considerations go beyond the scope of this investigation and will be investigated in future publications.

\subsection{Potential Applications}
\label{sec:Applications}

While the main purpose of this research was to see if and to what extent data can be binned across a spectrum without significant losses in the accuracy of a retrieval's predictions, there are other potential applications of this type of analysis. For instance, it would be interesting to try and asses the benefit of this work to future missions and alternative instrument modes available for the JWST.

In particular, this type of analysis could benefit the Ariel space mission since the mission will require efficient observation and analysis of atmospheres to probe the exoplanet population as broadly as possible within its lifetime. Ariel will use a tiered observing strategy with the survey level tier featuring binned spectra where appropriate. Understanding the maximum attainable information given a specific binning strategy (such as has already been considered in \cite{Alfnoor} and \cite{AlfnoorLowResInfo}) is important to the success of the mission.

With additional work, we hope that indicative binning paths may help to inform the necessary number of transits needed for accurate parameter estimations by identifying the photometric error that must be achieved. Providing a more holistic understanding of optimal observing strategies that takes into account optimal retrievability of planetary parameters given the minimally viable resolution and signal-to-noise regime of the observation, promises to increase the overall competitiveness of exoplanet transit and secondary eclipse observations on competitive instruments such as JWST and the ELTs. 

\section*{Acknowledgements}

This research received funding from the European Research Council (ERC) under the European Union's Horizon 2020 research and innovation programme (grant agreement n$^\circ$ 758892/ExoAI), and from the Science and Technology Facilities Council (STFC; grant n$^\circ$ ST/W00254X/1 and grant n$^\circ$ ST/W50788X/1).

%%%%%%%%%%%%%%%%%%%%%%%%%%%%%%%%%%%%%%%%%%%%%%%%%%
\section*{Data Availability}

Simulations generated for this study were based on the observational data presented in\,\cite{NIRSpecPRISM_Wasp39b} which can be accessed through the Mikulski Archive for Space Telescopes (MAST). Simulated data used in this investigation can be made available upon request.

%%%%%%%%%%%%%%%%%%%% REFERENCES %%%%%%%%%%%%%%%%%%

% The best way to enter references is to use BibTeX:

\bibliographystyle{mnras}
\bibliography{references} % if your bibtex file is called example.bib

\begin{thebibliography}{}
\makeatletter
\relax
\def\mn@urlcharsother{\let\do\@makeother \do\$\do\&\do\#\do\^\do\_\do\%\do\~}
\def\mn@doi{\begingroup\mn@urlcharsother \@ifnextchar [ {\mn@doi@} {\mn@doi@[]}}
\def\mn@doi@[#1]#2{\def\@tempa{#1}\ifx\@tempa\@empty \href {http://dx.doi.org/#2} {doi:#2}\else \href {http://dx.doi.org/#2} {#1}\fi \endgroup}
\def\mn@eprint#1#2{\mn@eprint@#1:#2::\@nil}
\def\mn@eprint@arXiv#1{\href {http://arxiv.org/abs/#1} {{\tt arXiv:#1}}}
\def\mn@eprint@dblp#1{\href {http://dblp.uni-trier.de/rec/bibtex/#1.xml} {dblp:#1}}
\def\mn@eprint@#1:#2:#3:#4\@nil{\def\@tempa {#1}\def\@tempb {#2}\def\@tempc {#3}\ifx \@tempc \@empty \let \@tempc \@tempb \let \@tempb \@tempa \fi \ifx \@tempb \@empty \def\@tempb {arXiv}\fi \@ifundefined {mn@eprint@\@tempb}{\@tempb:\@tempc}{\expandafter \expandafter \csname mn@eprint@\@tempb\endcsname \expandafter{\@tempc}}}

\bibitem[\protect\citeauthoryear{Ahrer et~al.,}{Ahrer et~al.}{2023a}]{JWST_NIRSpec_WASP39b_CO2}
Ahrer E.-M.,  et~al., 2023a, \mn@doi [Nature] {10.1038/s41586-022-05269-w}, 614, 649–652

\bibitem[\protect\citeauthoryear{Ahrer et~al.,}{Ahrer et~al.}{2023b}]{NIRCam_WASP39b}
Ahrer E.-M.,  et~al., 2023b, \mn@doi [Nature] {10.1038/s41586-022-05590-4}, 614, 653–658

\bibitem[\protect\citeauthoryear{Al-Refaie, Changeat, Waldmann  \& Tinetti}{Al-Refaie et~al.}{2021}]{TauREx3}
Al-Refaie A.~F.,  Changeat Q.,  Waldmann I.~P.,   Tinetti G.,  2021, \mn@doi [\apj] {10.3847/1538-4357/ac0252}, 917, 37

\bibitem[\protect\citeauthoryear{Al-Refaie, Changeat, Venot, Waldmann  \& Tinetti}{Al-Refaie et~al.}{2022}]{Al-Refaie2022_TauRexChemComparison}
Al-Refaie A.~F.,  Changeat Q.,  Venot O.,  Waldmann I.~P.,   Tinetti G.,  2022, \mn@doi [\apj] {10.3847/1538-4357/ac6dcd}, 932, 123

\bibitem[\protect\citeauthoryear{Alderson et~al.,}{Alderson et~al.}{2023}]{NIRSpecG395H_Wasp39b}
Alderson L.,  et~al., 2023, \mn@doi [Nature] {10.1038/s41586-022-05591-3}, 614, 664–669

\bibitem[\protect\citeauthoryear{Allard, Spiegelman, Leininger  \& Molliere}{Allard et~al.}{2019}]{Na_K}
Allard N.~F.,  Spiegelman F.,  Leininger T.,   Molliere P.,  2019, \mn@doi [\aap] {10.1051/0004-6361/201935593}, 628, A120

\bibitem[\protect\citeauthoryear{Arfaux \& Lavvas}{Arfaux \& Lavvas}{2024}]{arfaux2023_CloudsAndHaze_WASP39b}
Arfaux A.,  Lavvas P.,  2024, \mn@doi [\mnras] {10.1093/mnras/stae826}, 530, 482

\bibitem[\protect\citeauthoryear{Barman}{Barman}{2007}]{Barman2007_H2O}
Barman T.,  2007, \mn@doi [\apj] {10.1086/518736}, 661, L191

\bibitem[\protect\citeauthoryear{Barstow, Aigrain, Irwin, Kendrew  \& Fletcher}{Barstow et~al.}{2015}]{JWST_TransmissionSpecCapabilities}
Barstow J.~K.,  Aigrain S.,  Irwin P. G.~J.,  Kendrew S.,   Fletcher L.~N.,  2015, \mn@doi [\mnras] {10.1093/mnras/stv186}, 448, 2546–2561

\bibitem[\protect\citeauthoryear{Barstow, Aigrain, Irwin  \& Sing}{Barstow et~al.}{2016a}]{Barstow2016_PopStudy}
Barstow J.~K.,  Aigrain S.,  Irwin P. G.~J.,   Sing D.~K.,  2016a, \mn@doi [\apj] {10.3847/1538-4357/834/1/50}, 834, 50

\bibitem[\protect\citeauthoryear{Barstow, Aigrain, Irwin  \& Sing}{Barstow et~al.}{2016b}]{WASP39b_EqTemp}
Barstow J.~K.,  Aigrain S.,  Irwin P. G.~J.,   Sing D.~K.,  2016b, \mn@doi [\apj] {10.3847/1538-4357/834/1/50}, 834, 50

\bibitem[\protect\citeauthoryear{{Batalha}, {Kalirai}, {Lunine}, {Clampin}  \& {Lindler}}{{Batalha} et~al.}{2015}]{JWST_TransitingExoplanetSim}
{Batalha} N.,  {Kalirai} J.,  {Lunine} J.,  {Clampin} M.,   {Lindler} D.,  2015, \mn@doi [arXiv e-prints] {10.48550/arXiv.1507.02655}, \href {https://ui.adsabs.harvard.edu/abs/2015arXiv150702655B} {p. arXiv:1507.02655}

\bibitem[\protect\citeauthoryear{Bean, Abbot  \& Kempton}{Bean et~al.}{2017}]{Bean2017_ComparativePlanetology}
Bean J.~L.,  Abbot D.~S.,   Kempton E. M.-R.,  2017, \mn@doi [\apjl] {10.3847/2041-8213/aa738a}, 841, L24

\bibitem[\protect\citeauthoryear{Bean et~al.,}{Bean et~al.}{2018}]{JWST_ExoplanetERSPlan}
Bean J.~L.,  et~al., 2018, \mn@doi [\pasp] {10.1088/1538-3873/aadbf3}, 130, 114402

\bibitem[\protect\citeauthoryear{Bell et~al.,}{Bell et~al.}{2023}]{NIRCamWASP80b}
Bell T.~J.,  et~al., 2023, \mn@doi [Nature] {10.1038/s41586-023-06687-0}, 623, 709–712

\bibitem[\protect\citeauthoryear{Berta et~al.,}{Berta et~al.}{2012}]{Berta2012_GJ1214b_spec}
Berta Z.~K.,  et~al., 2012, \mn@doi [\apj] {10.1088/0004-637X/747/1/35}, 747, 35

\bibitem[\protect\citeauthoryear{Boldt-Christmas, Lesjak, Wehrhahn, Piskunov, Rains, Nortmann  \& Kochukhov}{Boldt-Christmas et~al.}{2023}]{HRS_SensitivityStudy}
Boldt-Christmas L.,  Lesjak F.,  Wehrhahn A.,  Piskunov N.,  Rains A.~D.,  Nortmann L.,   Kochukhov O.,  2023, \mn@doi [\aap] {10.48550/arXiv.2312.08320}

\bibitem[\protect\citeauthoryear{Bonomo et~al.,}{Bonomo et~al.}{2017}]{StellarParameters_Bonomo}
Bonomo A.~S.,  et~al., 2017, \mn@doi [\aap] {10.1051/0004-6361/201629882}, 602, A107

\bibitem[\protect\citeauthoryear{{Borysow, A.}}{{Borysow, A.}}{2002}]{H2_CIA_LowTemp}
{Borysow, A.} 2002, \mn@doi [A&A] {10.1051/0004-6361:20020555}, 390, 779

\bibitem[\protect\citeauthoryear{Borysow, Jørgensen  \& Fu}{Borysow et~al.}{2001}]{H2_CIA_HighTemp}
Borysow A.,  Jørgensen U.~G.,   Fu Y.,  2001, \mn@doi [Journal of Quantitative Spectroscopy and Radiative Transfer] {https://doi.org/10.1016/S0022-4073(00)00023-6}, 68, 235

\bibitem[\protect\citeauthoryear{{Buchner, J.} et~al.,}{{Buchner, J.} et~al.}{2014}]{Buchner2014_PyMultinest}
{Buchner, J.} et~al., 2014, \mn@doi [A&A] {10.1051/0004-6361/201322971}, 564, A125

\bibitem[\protect\citeauthoryear{Changeat, Edwards, Waldmann  \& Tinetti}{Changeat et~al.}{2019}]{Changeat_TwoLayerChem}
Changeat Q.,  Edwards B.,  Waldmann I.~P.,   Tinetti G.,  2019, \mn@doi [\apj] {10.3847/1538-4357/ab4a14}, 886, 39

\bibitem[\protect\citeauthoryear{Changeat, Al-Refaie, Mugnai, Edwards, Waldmann, Pascale  \& Tinetti}{Changeat et~al.}{2020}]{Alfnoor}
Changeat Q.,  Al-Refaie A.,  Mugnai L.~V.,  Edwards B.,  Waldmann I.~P.,  Pascale E.,   Tinetti G.,  2020, \mn@doi [\aj] {10.3847/1538-3881/ab9a53}, 160, 80

\bibitem[\protect\citeauthoryear{Changeat et~al.,}{Changeat et~al.}{2022}]{changeat2022_PopStudy}
Changeat Q.,  et~al., 2022, \mn@doi [\apjs] {10.3847/1538-4365/ac5cc2}, 260, 3

\bibitem[\protect\citeauthoryear{Chubb et~al.,}{Chubb et~al.}{2021}]{ExoMol3}
Chubb K.~L.,  et~al., 2021, \mn@doi [\aap] {10.1051/0004-6361/202038350}, 646, A21

\bibitem[\protect\citeauthoryear{Crouzet, Mccullough, Burke  \& Long}{Crouzet et~al.}{2012}]{Crouzet2012_XO2b_spec}
Crouzet N.,  Mccullough P.~R.,  Burke C.,   Long D.,  2012, \mn@doi [\apj] {10.1088/0004-637x/761/1/7}, 761, 7

\bibitem[\protect\citeauthoryear{{Deming} et~al.,}{{Deming} et~al.}{2013}]{Deming2013_HD209458b_spec}
{Deming} D.,  et~al., 2013, \mn@doi [\apj] {10.1088/0004-637X/774/2/95}, \href {https://ui.adsabs.harvard.edu/abs/2013ApJ...774...95D} {774, 95}

\bibitem[\protect\citeauthoryear{Edwards et~al.,}{Edwards et~al.}{2023}]{HST_EdwardsPopStudy}
Edwards B.,  et~al., 2023, \mn@doi [\apjs] {10.3847/1538-4365/ac9f1a}, 269, 31

\bibitem[\protect\citeauthoryear{{Faedi} et~al.,}{{Faedi} et~al.}{2011}]{Wasp39b_Discovery}
{Faedi} F.,  et~al., 2011, \mn@doi [\aap] {10.1051/0004-6361/201116671}, \href {https://ui.adsabs.harvard.edu/abs/2011A&A...531A..40F} {531, A40}

\bibitem[\protect\citeauthoryear{Feinstein et~al.,}{Feinstein et~al.}{2023}]{NIRISS_WASP39b}
Feinstein A.~D.,  et~al., 2023, \mn@doi [Nature] {10.1038/s41586-022-05674-1}, 614, 670–675

\bibitem[\protect\citeauthoryear{Feng, Robinson, Fortney, Lupu, Marley, Lewis, Macintosh  \& Line}{Feng et~al.}{2018}]{Feng_NoNeedToScatter}
Feng Y.~K.,  Robinson T.~D.,  Fortney J.~J.,  Lupu R.~E.,  Marley M.~S.,  Lewis N.~K.,  Macintosh B.,   Line M.~R.,  2018, \mn@doi [\aj] {10.3847/1538-3881/aab95c}, 155, 200

\bibitem[\protect\citeauthoryear{{Feroz} \& {Hobson}}{{Feroz} \& {Hobson}}{2008}]{MultiNest_1}
{Feroz} F.,  {Hobson} M.~P.,  2008, \mn@doi [\mnras] {10.1111/j.1365-2966.2007.12353.x}, \href {https://ui.adsabs.harvard.edu/abs/2008MNRAS.384..449F} {384, 449}

\bibitem[\protect\citeauthoryear{{Feroz}, {Hobson}  \& {Bridges}}{{Feroz} et~al.}{2009}]{MultiNest_2}
{Feroz} F.,  {Hobson} M.~P.,   {Bridges} M.,  2009, \mn@doi [\mnras] {10.1111/j.1365-2966.2009.14548.x}, \href {https://ui.adsabs.harvard.edu/abs/2009MNRAS.398.1601F} {398, 1601}

\bibitem[\protect\citeauthoryear{{Feroz}, {Hobson}, {Cameron}  \& {Pettitt}}{{Feroz} et~al.}{2019}]{MultiNest_3}
{Feroz} F.,  {Hobson} M.~P.,  {Cameron} E.,   {Pettitt} A.~N.,  2019, \mn@doi [The Open Journal of Astrophysics] {10.21105/astro.1306.2144}, \href {https://ui.adsabs.harvard.edu/abs/2019OJAp....2E..10F} {2, 10}

\bibitem[\protect\citeauthoryear{Fischer et~al.,}{Fischer et~al.}{2016}]{HST_STIS_Wasp39b}
Fischer P.~D.,  et~al., 2016, \mn@doi [\apj] {10.3847/0004-637X/827/1/19}, 827, 19

\bibitem[\protect\citeauthoryear{Fisher \& Heng}{Fisher \& Heng}{2018}]{Fisher2018_PopStudy}
Fisher C.,  Heng K.,  2018, \mn@doi [\mnras] {10.1093/mnras/sty2550}, 481, 4698

\bibitem[\protect\citeauthoryear{Greene, Line, Montero, Fortney, Lustig-Yaeger  \& Luther}{Greene et~al.}{2016}]{JWST_TransmissionSpecOverview}
Greene T.~P.,  Line M.~R.,  Montero C.,  Fortney J.~J.,  Lustig-Yaeger J.,   Luther K.,  2016, \mn@doi [\apj] {10.3847/0004-637X/817/1/17}, 817, 17

\bibitem[\protect\citeauthoryear{Guillot}{Guillot}{2010}]{GuillotTempProfile}
Guillot T.,  2010, \mn@doi [\aap] {10.1051/0004-6361/200913396}, 520, A27

\bibitem[\protect\citeauthoryear{Heng, Mendonça  \& Lee}{Heng et~al.}{2014}]{HengTwoStreamRadTransfer}
Heng K.,  Mendonça J.~M.,   Lee J.-M.,  2014, \mn@doi [\apjs] {10.1088/0067-0049/215/1/4}, 215, 4

\bibitem[\protect\citeauthoryear{Irwin et~al.,}{Irwin et~al.}{2008}]{NEMESIS}
Irwin P.,  et~al., 2008, \mn@doi [Journal of Quantitative Spectroscopy and Radiative Transfer] {https://doi.org/10.1016/j.jqsrt.2007.11.006}, 109, 1136

\bibitem[\protect\citeauthoryear{Jakobsen et~al.,}{Jakobsen et~al.}{2022}]{NIRSpecOverview}
Jakobsen P.,  et~al., 2022, \mn@doi [\aap] {10.1051/0004-6361/202142663}, 661, A80

\bibitem[\protect\citeauthoryear{Kipping}{Kipping}{2010}]{TemporalBinning_Transit}
Kipping D.~M.,  2010, \mn@doi [\mnras] {10.1111/j.1365-2966.2010.17242.x}, 408, 1758

\bibitem[\protect\citeauthoryear{{Kirk}, {L{\'o}pez-Morales}, {Wheatley}, {Weaver}, {Skillen}, {Louden}, {McCormac}  \& {Espinoza}}{{Kirk} et~al.}{2019}]{LRG_BEASTS_WASP39b}
{Kirk} J.,  {L{\'o}pez-Morales} M.,  {Wheatley} P.~J.,  {Weaver} I.~C.,  {Skillen} I.,  {Louden} T.,  {McCormac} J.,   {Espinoza} N.,  2019, \mn@doi [\aj] {10.3847/1538-3881/ab397d}, \href {https://ui.adsabs.harvard.edu/abs/2019AJ....158..144K} {158, 144}

\bibitem[\protect\citeauthoryear{Kreidberg et~al.,}{Kreidberg et~al.}{2014a}]{Kreidberg2014_CloudsSuperEarth}
Kreidberg L.,  et~al., 2014a, \mn@doi [Nature] {10.1038/nature12888}, 505, 69–72

\bibitem[\protect\citeauthoryear{{Kreidberg} et~al.,}{{Kreidberg} et~al.}{2014b}]{Kreidberg2014_wasp43b_spec}
{Kreidberg} L.,  et~al., 2014b, \mn@doi [\apjl] {10.1088/2041-8205/793/2/L27}, \href {https://ui.adsabs.harvard.edu/abs/2014ApJ...793L..27K} {793, L27}

\bibitem[\protect\citeauthoryear{{Kreidberg} et~al.,}{{Kreidberg} et~al.}{2015}]{Kreidberg2015_wasp12b_spec}
{Kreidberg} L.,  et~al., 2015, \mn@doi [\apj] {10.1088/0004-637X/814/1/66}, \href {https://ui.adsabs.harvard.edu/abs/2015ApJ...814...66K} {814, 66}

\bibitem[\protect\citeauthoryear{Li, Gordon, Rothman, Tan, Hu, Kassi, Campargue  \& Medvedev}{Li et~al.}{2015}]{CO}
Li G.,  Gordon I.~E.,  Rothman L.~S.,  Tan Y.,  Hu S.-M.,  Kassi S.,  Campargue A.,   Medvedev E.~S.,  2015, \mn@doi [\apjs] {10.1088/0067-0049/216/1/15}, 216, 15

\bibitem[\protect\citeauthoryear{Line et~al.,}{Line et~al.}{2013}]{CHIMERA}
Line M.~R.,  et~al., 2013, \mn@doi [\apj] {10.1088/0004-637X/775/2/137}, 775, 137

\bibitem[\protect\citeauthoryear{Lustig-Yaeger et~al.,}{Lustig-Yaeger et~al.}{2023}]{JWST_EarthSizedExoplanSpec}
Lustig-Yaeger J.,  et~al., 2023, \mn@doi [Nature Astronomy] {10.1038/s41550-023-02064-z}

\bibitem[\protect\citeauthoryear{MacDonald \& Madhusudhan}{MacDonald \& Madhusudhan}{2017}]{1MacDonald2017_PatchyClouds_HD209}
MacDonald R.~J.,  Madhusudhan N.,  2017, \mn@doi [\mnras] {10.1093/mnras/stx804}, 469, 1979

\bibitem[\protect\citeauthoryear{MacDonald, Goyal  \& Lewis}{MacDonald et~al.}{2020}]{EffTempErrors}
MacDonald R.~J.,  Goyal J.~M.,   Lewis N.~K.,  2020, \mn@doi [\apjl] {10.3847/2041-8213/ab8238}, 893, L43

\bibitem[\protect\citeauthoryear{Macdonald \& Batalha}{Macdonald \& Batalha}{2023}]{ListRetrievalCodes_2023}
Macdonald R.~J.,  Batalha N.~E.,  2023, \mn@doi [Research Notes of the AAS] {10.3847/2515-5172/acc46a}, 7, 54

\bibitem[\protect\citeauthoryear{{Maciejewski} et~al.,}{{Maciejewski} et~al.}{2016}]{WASP39bParameters_Maciejewski}
{Maciejewski} G.,  et~al., 2016, \mn@doi [\actaa] {10.48550/arXiv.1603.03268}, \href {https://ui.adsabs.harvard.edu/abs/2016AcA....66...55M} {66, 55}

\bibitem[\protect\citeauthoryear{Madhusudhan}{Madhusudhan}{2018}]{RetrievalSummaryChap}
Madhusudhan N.,  2018, Atmospheric Retrieval of Exoplanets.
p. 2153–2182, \mn@doi{10.1007/978-3-319-55333-7_104}

\bibitem[\protect\citeauthoryear{{Madhusudhan} \& {Seager}}{{Madhusudhan} \& {Seager}}{2009}]{FirstRetrievals}
{Madhusudhan} N.,  {Seager} S.,  2009, \mn@doi [\apj] {10.1088/0004-637X/707/1/24}, \href {https://ui.adsabs.harvard.edu/abs/2009ApJ...707...24M} {707, 24}

\bibitem[\protect\citeauthoryear{Malik et~al.,}{Malik et~al.}{2017}]{HELIOS_TwoStreamRadTransfer}
Malik M.,  et~al., 2017, \mn@doi [\aj] {10.3847/1538-3881/153/2/56}, 153, 56

\bibitem[\protect\citeauthoryear{Mollière, Wardenier, van Boekel, Henning, Molaverdikhani  \& Snellen}{Mollière et~al.}{2019}]{petitRADTRANS}
Mollière P.,  Wardenier J.~P.,  van Boekel R.,  Henning T.,  Molaverdikhani K.,   Snellen I. A.~G.,  2019, \mn@doi [\aap] {10.1051/0004-6361/201935470}, 627, A67

\bibitem[\protect\citeauthoryear{Morello, Dyrek  \& Changeat}{Morello et~al.}{2022}]{TemporalBinning_PhaseCurves}
Morello G.,  Dyrek A.,   Changeat Q.,  2022, \mn@doi [\mnras] {10.1093/mnras/stac2828}, 517, 2151–2164

\bibitem[\protect\citeauthoryear{Morley, Kreidberg, Rustamkulov, Robinson  \& Fortney}{Morley et~al.}{2017}]{JWST_ObservingEarthSizedExoplanets}
Morley C.~V.,  Kreidberg L.,  Rustamkulov Z.,  Robinson T.,   Fortney J.~J.,  2017, \mn@doi [\apj] {10.3847/1538-4357/aa927b}, 850, 121

\bibitem[\protect\citeauthoryear{Mugnai, Al-Refaie, Bocchieri, Changeat, Pascale  \& Tinetti}{Mugnai et~al.}{2021}]{AlfnoorLowResInfo}
Mugnai L.~V.,  Al-Refaie A.,  Bocchieri A.,  Changeat Q.,  Pascale E.,   Tinetti G.,  2021, \mn@doi [\aj] {10.3847/1538-3881/ac2e92}, 162, 288

\bibitem[\protect\citeauthoryear{{Nikolov}, {Sing}, {Gibson}, {Fortney}, {Evans}, {Barstow}, {Kataria}  \& {Wilson}}{{Nikolov} et~al.}{2016}]{VLT_WASP39b}
{Nikolov} N.,  {Sing} D.~K.,  {Gibson} N.~P.,  {Fortney} J.~J.,  {Evans} T.~M.,  {Barstow} J.~K.,  {Kataria} T.,   {Wilson} P.~A.,  2016, \mn@doi [\apj] {10.3847/0004-637X/832/2/191}, \href {https://ui.adsabs.harvard.edu/abs/2016ApJ...832..191N} {832, 191}

\bibitem[\protect\citeauthoryear{Pinhas, Madhusudhan, Gandhi  \& MacDonald}{Pinhas et~al.}{2018}]{Pinhas2018_HotJupH2OSurvey}
Pinhas A.,  Madhusudhan N.,  Gandhi S.,   MacDonald R.,  2018, \mn@doi [\mnras] {10.1093/mnras/sty2544}, 482, 1485

\bibitem[\protect\citeauthoryear{Pluriel, Zingales, Leconte  \& Parmentier}{Pluriel et~al.}{2020}]{ChemistryBias}
Pluriel W.,  Zingales T.,  Leconte J.,   Parmentier V.,  2020, \mn@doi [\aap] {10.1051/0004-6361/202037678}, 636, A66

\bibitem[\protect\citeauthoryear{Polyansky, Kyuberis, Zobov, Tennyson, Yurchenko  \& Lodi}{Polyansky et~al.}{2018}]{H2O}
Polyansky O.~L.,  Kyuberis A.~A.,  Zobov N.~F.,  Tennyson J.,  Yurchenko S.~N.,   Lodi L.,  2018, \mn@doi [\mnras] {10.1093/mnras/sty1877}, 480, 2597

\bibitem[\protect\citeauthoryear{Richard et~al.,}{Richard et~al.}{2012}]{RICHARD2012_HITRAN_CIA}
Richard C.,  et~al., 2012, \mn@doi [Journal of Quantitative Spectroscopy and Radiative Transfer] {https://doi.org/10.1016/j.jqsrt.2011.11.004}, 113, 1276

\bibitem[\protect\citeauthoryear{Rothman et~al.,}{Rothman et~al.}{2013}]{CIA_Hitran_2013}
Rothman L.,  et~al., 2013, \mn@doi [Journal of Quantitative Spectroscopy and Radiative Transfer] {https://doi.org/10.1016/j.jqsrt.2013.07.002}, 130, 4

\bibitem[\protect\citeauthoryear{Rustamkulov, Sing, Liu  \& Wang}{Rustamkulov et~al.}{2022}]{FIREFLy}
Rustamkulov Z.,  Sing D.~K.,  Liu R.,   Wang A.,  2022, \mn@doi [\apjl] {10.3847/2041-8213/ac5b6f}, 928, L7

\bibitem[\protect\citeauthoryear{Rustamkulov et~al.,}{Rustamkulov et~al.}{2023}]{NIRSpecPRISM_Wasp39b}
Rustamkulov Z.,  et~al., 2023, \mn@doi [Nature] {10.1038/s41586-022-05677-y}, 614, 659–663

\bibitem[\protect\citeauthoryear{Saba et~al.,}{Saba et~al.}{2022}]{Saba2022_wasp17b_spec}
Saba A.,  et~al., 2022, \mn@doi [\aj] {10.3847/1538-3881/ac6c01}, 164, 2

\bibitem[\protect\citeauthoryear{Sing et~al.,}{Sing et~al.}{2016}]{Sing2016_TenHotJupiters}
Sing D.~K.,  et~al., 2016, \mn@doi [Nature] {10.1038/nature16068}, 529, 59–62

\bibitem[\protect\citeauthoryear{Skaf et~al.,}{Skaf et~al.}{2020}]{Eff_Eqm_TempRetrievals}
Skaf N.,  et~al., 2020, \mn@doi [\aj] {10.3847/1538-3881/ab94a3}, 160, 109

\bibitem[\protect\citeauthoryear{Skilling}{Skilling}{2004}]{Skilling_NestedSamp}
Skilling J.,  2004, in 8TH BRUNEI INTERNATIONAL CONFERENCE ON ENGINEERING AND TECHNOLOGY 2021. 8TH BRUNEI INTERNATIONAL CONFERENCE ON ENGINEERING AND TECHNOLOGY 2021, \mn@doi{10.1063/1.1835238}

\bibitem[\protect\citeauthoryear{Swain, Vasisht  \& Tinetti}{Swain et~al.}{2008}]{Swain2008_CH4}
Swain M.~R.,  Vasisht G.,   Tinetti G.,  2008, \mn@doi [Nature] {10.1038/nature06823}, 452, 329–331

\bibitem[\protect\citeauthoryear{Swain, Vasisht, Tinetti, Bouwman, Chen, Yung, Deming  \& Deroo}{Swain et~al.}{2009}]{Swain2009_HD189733b_spec}
Swain M.~R.,  Vasisht G.,  Tinetti G.,  Bouwman J.,  Chen P.,  Yung Y.,  Deming D.,   Deroo P.,  2009, \mn@doi [\apj] {10.1088/0004-637x/690/2/l114}, 690, L114–L117

\bibitem[\protect\citeauthoryear{Swain, Line  \& Deroo}{Swain et~al.}{2014}]{Swain2014_189733b_spec}
Swain M.~R.,  Line M.~R.,   Deroo P.,  2014, \mn@doi [\apj] {10.1088/0004-637X/784/2/133}, 784, 133

\bibitem[\protect\citeauthoryear{{Tennyson} \& {Yurchenko}}{{Tennyson} \& {Yurchenko}}{2012}]{ExoMol1}
{Tennyson} J.,  {Yurchenko} S.~N.,  2012, \mn@doi [\mnras] {10.1111/j.1365-2966.2012.21440.x}, \href {https://ui.adsabs.harvard.edu/abs/2012MNRAS.425...21T} {425, 21}

\bibitem[\protect\citeauthoryear{{Tennyson} et~al.,}{{Tennyson} et~al.}{2016}]{ExoMol2}
{Tennyson} J.,  et~al., 2016, \mn@doi [Journal of Molecular Spectroscopy] {10.1016/j.jms.2016.05.002}, \href {https://ui.adsabs.harvard.edu/abs/2016JMoSp.327...73T} {327, 73}

\bibitem[\protect\citeauthoryear{Tinetti et~al.,}{Tinetti et~al.}{2007}]{Tinetti2007_H2O}
Tinetti G.,  et~al., 2007, \mn@doi [Nature] {10.1038/nature06002}, 448, 169–171

\bibitem[\protect\citeauthoryear{Tinetti et~al.,}{Tinetti et~al.}{2018}]{ARIEL_Overview}
Tinetti G.,  et~al., 2018, \mn@doi [Experimental Astronomy] {10.1007/s10686-018-9598-x}, 46, 135–209

\bibitem[\protect\citeauthoryear{Tsai et~al.,}{Tsai et~al.}{2023}]{photochem_wasp39b}
Tsai S.-M.,  et~al., 2023, \mn@doi [Nature] {10.1038/s41586-023-05902-2}, 617, 483–487

\bibitem[\protect\citeauthoryear{Tsiaras et~al.,}{Tsiaras et~al.}{2018}]{HST_TsiarasPopStudy}
Tsiaras A.,  et~al., 2018, \mn@doi [\aj] {10.3847/1538-3881/aaaf75}, 155, 156

\bibitem[\protect\citeauthoryear{{Tsiaras}, {Waldmann}, {Tinetti}, {Tennyson}  \& {Yurchenko}}{{Tsiaras} et~al.}{2019}]{Tsiaras2019_K218b_spec}
{Tsiaras} A.,  {Waldmann} I.~P.,  {Tinetti} G.,  {Tennyson} J.,   {Yurchenko} S.~N.,  2019, \mn@doi [Nature Astronomy] {10.1038/s41550-019-0878-9}, \href {https://ui.adsabs.harvard.edu/abs/2019NatAs...3.1086T} {3, 1086}

\bibitem[\protect\citeauthoryear{Underwood, Tennyson, Yurchenko, Huang, Schwenke, Lee, Clausen  \& Fateev}{Underwood et~al.}{2016}]{SO2}
Underwood D.~S.,  Tennyson J.,  Yurchenko S.~N.,  Huang X.,  Schwenke D.~W.,  Lee T.~J.,  Clausen S.,   Fateev A.,  2016, \mn@doi [\mnras] {10.1093/mnras/stw849}, 459, 3890

\bibitem[\protect\citeauthoryear{{Wakeford} et~al.,}{{Wakeford} et~al.}{2018}]{HST_WFC3_WASP39b}
{Wakeford} H.~R.,  et~al., 2018, \mn@doi [\aj] {10.3847/1538-3881/aa9e4e}, \href {https://ui.adsabs.harvard.edu/abs/2018AJ....155...29W} {155, 29}

\bibitem[\protect\citeauthoryear{{Waldmann}, {Tinetti}, {Rocchetto}, {Barton}, {Yurchenko}  \& {Tennyson}}{{Waldmann} et~al.}{2015a}]{TauREx1}
{Waldmann} I.~P.,  {Tinetti} G.,  {Rocchetto} M.,  {Barton} E.~J.,  {Yurchenko} S.~N.,   {Tennyson} J.,  2015a, \mn@doi [\apj] {10.1088/0004-637X/802/2/107}, \href {https://ui.adsabs.harvard.edu/abs/2015ApJ...802..107W} {802, 107}

\bibitem[\protect\citeauthoryear{{Waldmann}, {Rocchetto}, {Tinetti}, {Barton}, {Yurchenko}  \& {Tennyson}}{{Waldmann} et~al.}{2015b}]{TauREx2}
{Waldmann} I.~P.,  {Rocchetto} M.,  {Tinetti} G.,  {Barton} E.~J.,  {Yurchenko} S.~N.,   {Tennyson} J.,  2015b, \mn@doi [\apj] {10.1088/0004-637X/813/1/13}, \href {https://ui.adsabs.harvard.edu/abs/2015ApJ...813...13W} {813, 13}

\bibitem[\protect\citeauthoryear{Yurchenko, Mellor, Freedman  \& Tennyson}{Yurchenko et~al.}{2020}]{CO2}
Yurchenko S.~N.,  Mellor T.~M.,  Freedman R.~S.,   Tennyson J.,  2020, \mn@doi [\mnras] {10.1093/mnras/staa1874}, 496, 5282

\makeatother
\end{thebibliography}

%%%%%%%%%%%%%%%%%%%%%%%%%%%%%%%%%%%%%%%%%%%%%%%%%%

%%%%%%%%%%%%%%%%% APPENDICES %%%%%%%%%%%%%%%%%%%%%

\appendix

\section{Description of Appendices}
\label{Appendix:AdditionalPlots}

Additional figures are included in these appendices. In most cases these represent either additional tests of the reliability or stability of our results under the scrutiny of further investigation or additional grids which follow similar trends to those mentioned in the main text possibly for a different set up (e.g. varying cloud levels).

Section \ref{App:PostWidthGrids} provides the maps of the width of the posterior distribution for each parameter in the low and no cloud cases. Similarly, section \ref{App:HighResMaps} provides supplementary grids for the bias of retrievals on the high resolution end of the retrieval grid in each cloud case. \ref{App:CornerPlots} provides three overlaid corner plots to provide further evidence for the discussion of figure~\ref{fig:SpecContributions} in the main text. In~\ref{App:BinFactor} further plots considering retrievals on spectra binned from the ERS wavelength grid are presented and \ref{App:BinFactor_shift} shows similar plots for the retrievals on a shifted wavelength grid. \ref{App:Complexity} shows our investigation of the effects of model complexity on the retrieval bias while \ref{App:Repeat} and \ref{App:Perturb} show plots considering the difference in the retrieval bias under repeat retrievals with or without additional scatter on the spectra. 

\subsection{Posterior width grids}
\label{App:PostWidthGrids}

\begin{figure*}
    \includegraphics[width=\textwidth]{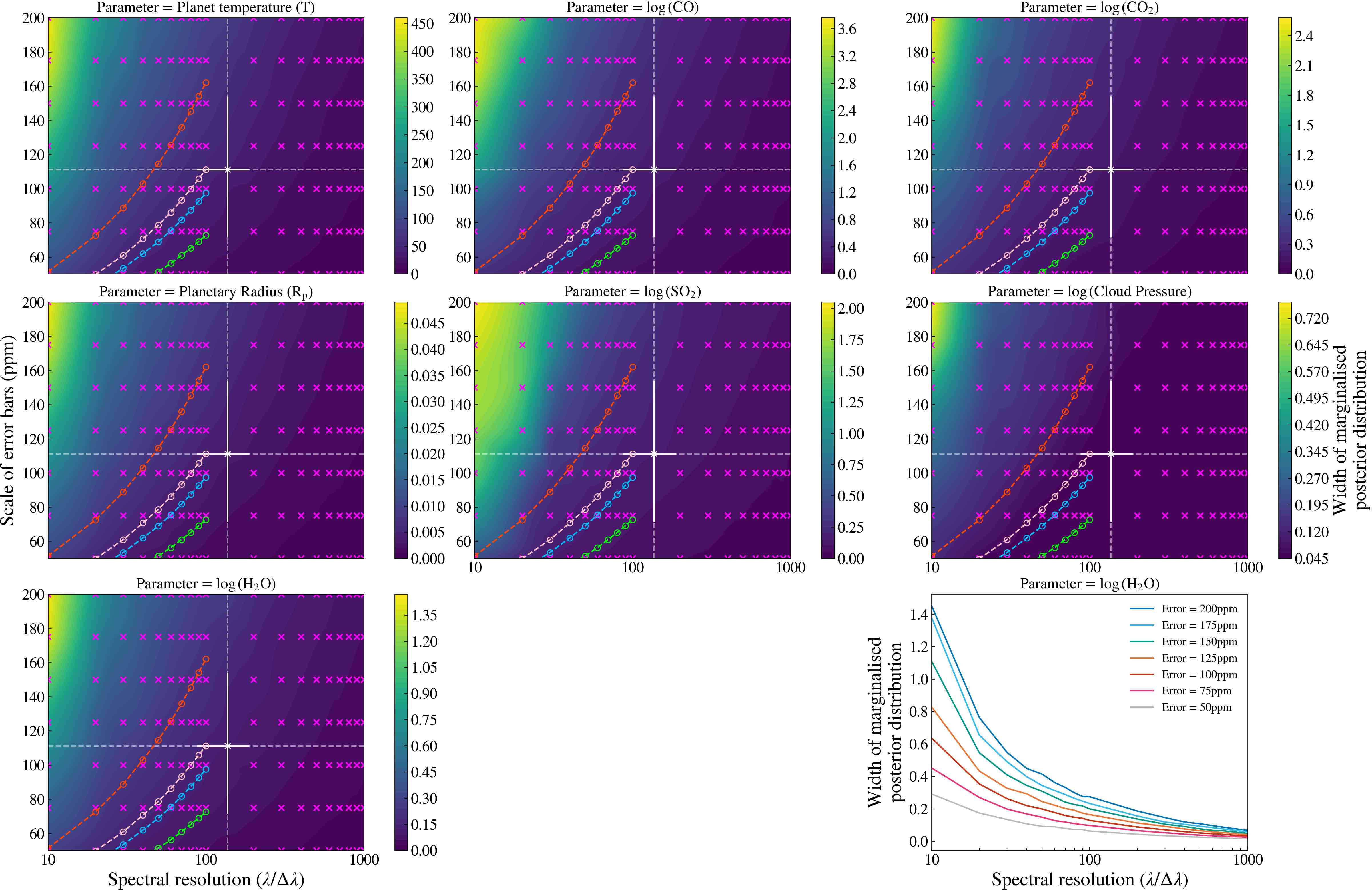}
    \caption{Posterior width maps for the case of a low cloud deck. In each subplot (labelled by which parameter is represented) the pink crosses mark the position of data from retrievals. Interpolation has been used to fill in the rest of the grid and generate a map. A log scale is used for the spectral resolution (x axis) since these data are spaced by 10 below R~=~100 and by 100 between R~=~100 and R~=~500. The colour bar associated with each map describes the width of the posterior distribution obtained in retrievals for the given parameter. The white arrow marks the approximate position of the JWST ERS data in the map (including error bars set by the 25\textsuperscript{th} and 75\textsuperscript{th} percentiles of the variation across the observed spectrum in resolution and error bar size). The green dashed lines plot potential binning paths under the assumption that each bin (before binning) contains an equal number of counts. The pink dashed line plots the potential binning path from the approximate position of the ERS data point under the same assumptions. The lower right plot in the grid shows the same data as the lower left but as a series of lines.}
    \label{fig:LowCloudWidthMap}
\end{figure*}

\begin{figure*}
    \includegraphics[width=\textwidth]{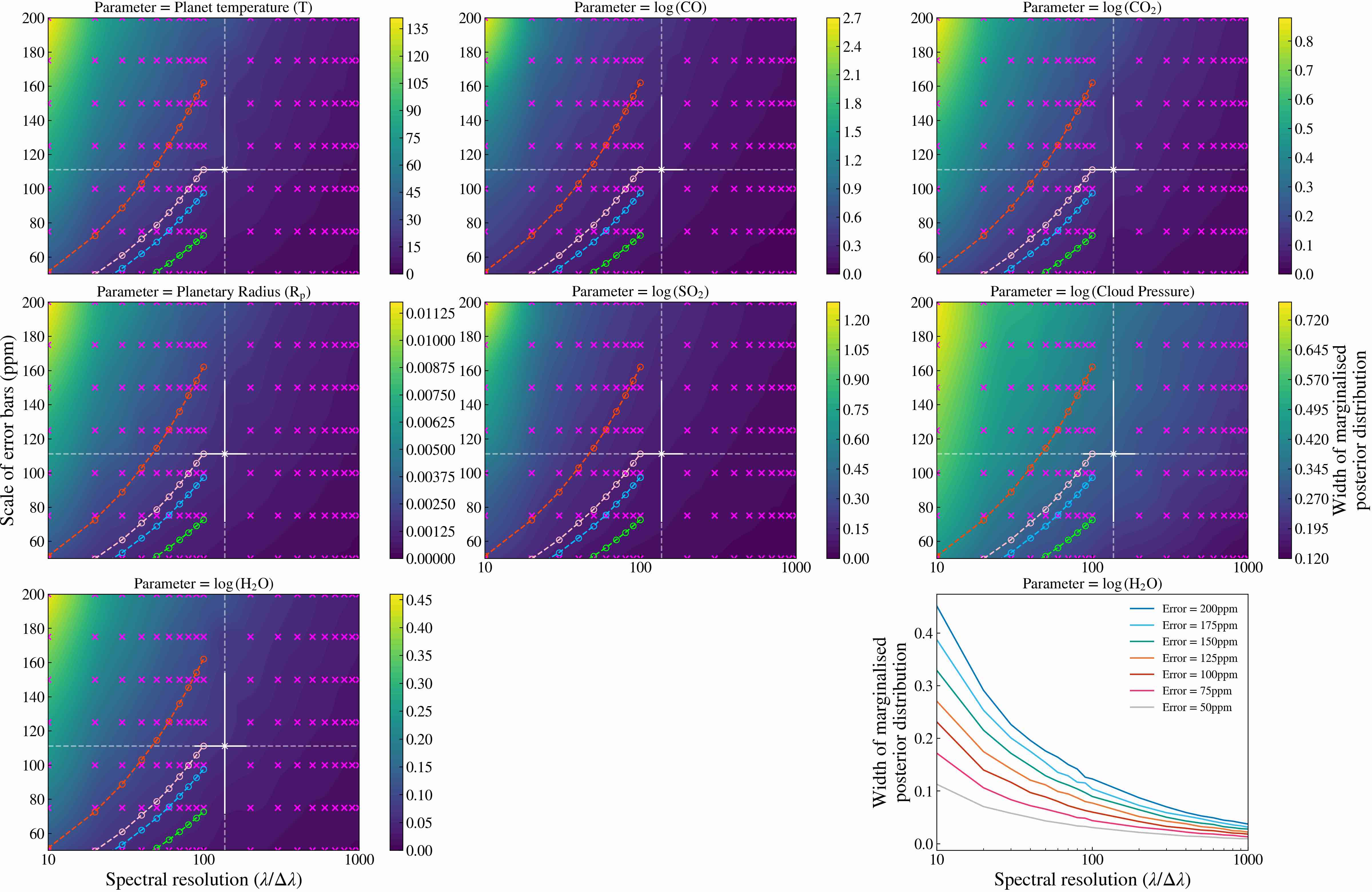}
    \caption{The same as figure~\ref{fig:LowCloudWidthMap} except in the case of no clouds rather than a low cloud deck.}
    \label{fig:NoCloudWidthMap}
\end{figure*}

Figures \ref{fig:LowCloudWidthMap} and \ref{fig:NoCloudWidthMap} show the maps of the posterior distribution widths plotted for the cases of low and no clouds respectively. These plots are formatted in the same way as figure \ref{fig:HighCloudWidthMap} from the main text. They are included here for completeness but only show the expected trend that the width of the posterior distribution decreased as resolution increased and photometric error decreased.

\subsection{High resolution bias maps}
\label{App:HighResMaps}

\begin{figure*}
    \includegraphics[width=\textwidth]{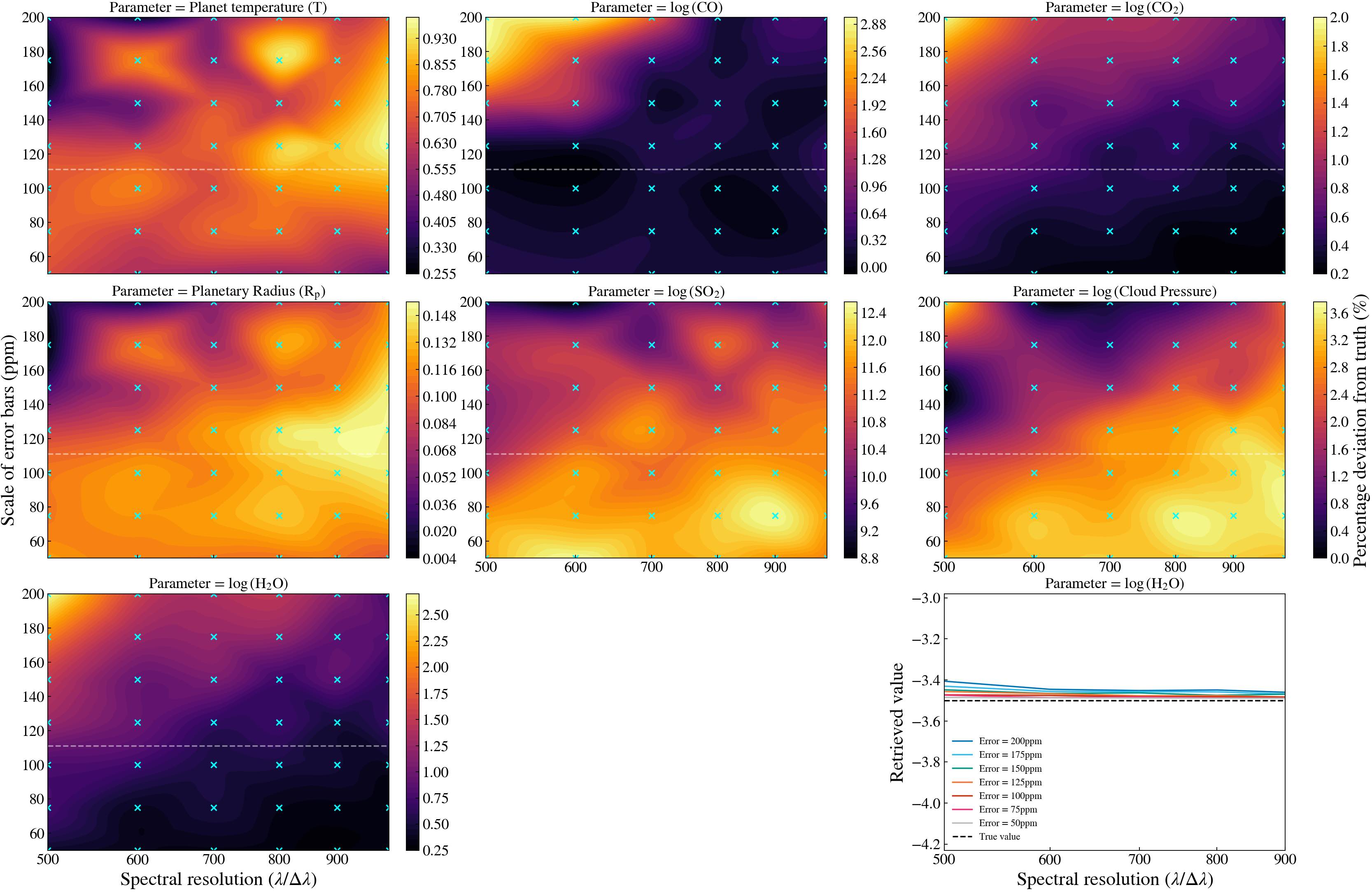}
    \caption{Sensitivity maps for the case of a high cloud deck included in the simulation. In each subplot (labelled by which parameter is represented) the blue crosses mark the position of data from retrievals. Interpolation has been used to fill in the rest of the grid and generate a map. A log scale is used for the spectral resolution (x axis). The colour bar associated with each map describes the percentage deviation of the value obtained in retrievals from the true value of that parameter (input to the simulation). Note that the faint dashed line indicates the position of the ERS data point in the parameter space. The point itself is not visible since it lies outside of the resolution range of these plots.}
    \label{fig:HighCloudSenMap_HighRes}
\end{figure*}

\begin{figure*}
    \includegraphics[width=\textwidth]{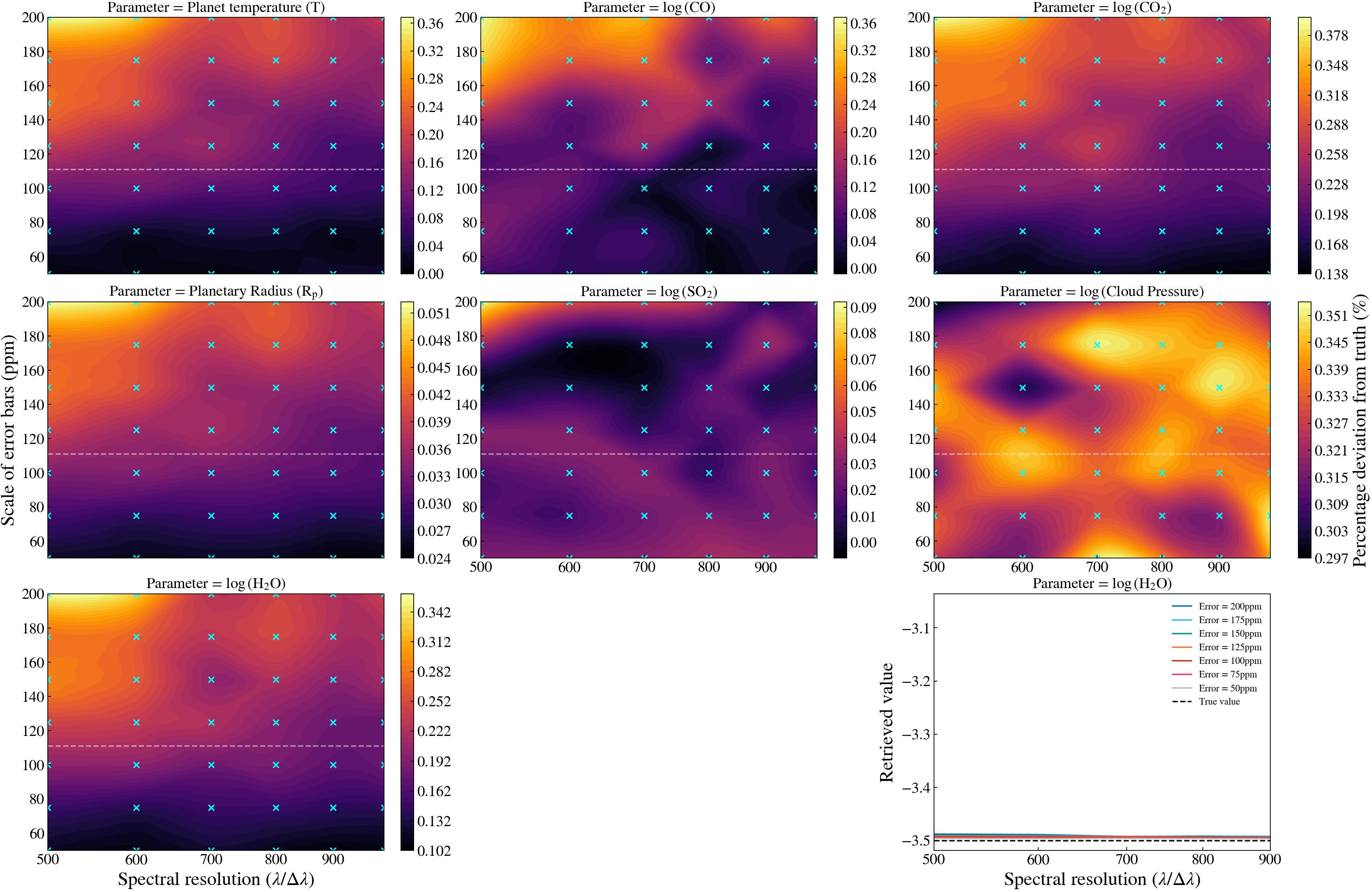}
    \caption{Same as figure~\ref{fig:HighCloudSenMap_HighRes} except that here we show data from retrievals on the low cloud spectra.}
    \label{fig:LowCloudSenMap_HighRes}
\end{figure*}

\begin{figure*}
    \includegraphics[width=\textwidth]{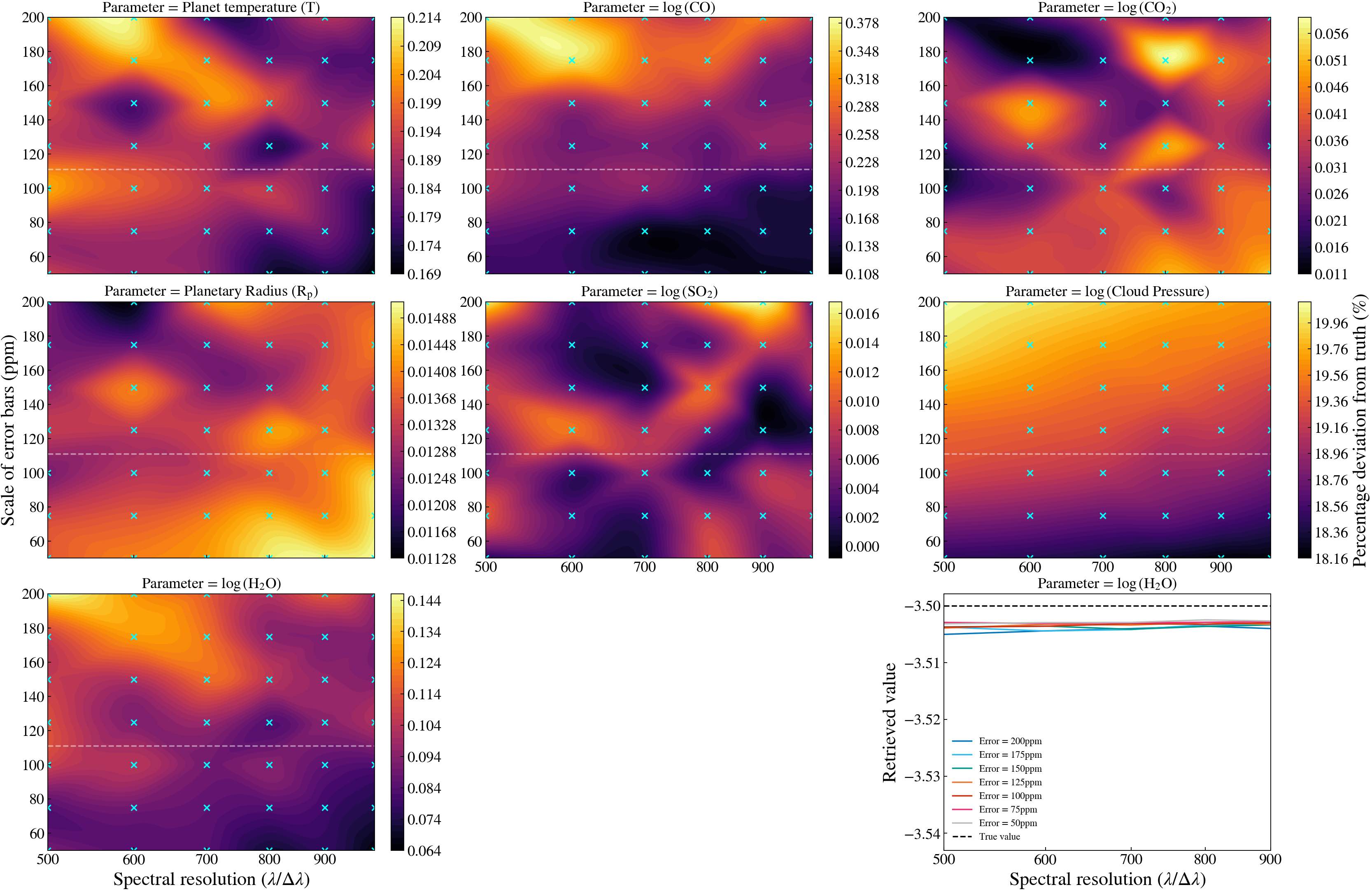}
    \caption{Same as figure~\ref{fig:HighCloudSenMap_HighRes} except that here we show data from retrievals on the cloud free spectra.}
    \label{fig:NoCloudSenMap_HighRes}
\end{figure*}

In figures \ref{fig:HighCloudSenMap_HighRes}, \ref{fig:LowCloudSenMap_HighRes} and \ref{fig:NoCloudSenMap_HighRes} we plot the bias maps for the high resolution end of the retrieval grid in the cases of a high cloud deck, a low cloud deck and no clouds respectively. These should be seen to compliment figures \ref{fig:HighCloudSenMap}, \ref{fig:LowCloudSenMap} and \ref{fig:NoCloudSenMap} in the main text which show the low resolution end of the grid (up to resolutions of R\,=\,500). Less deviation is seen at resolutions between 500 and 1000 so they are left to the appendices rather than the main text.

\subsection{Retrieval corner plots}
\label{App:CornerPlots}

\begin{figure*}
    \includegraphics[width=\textwidth]{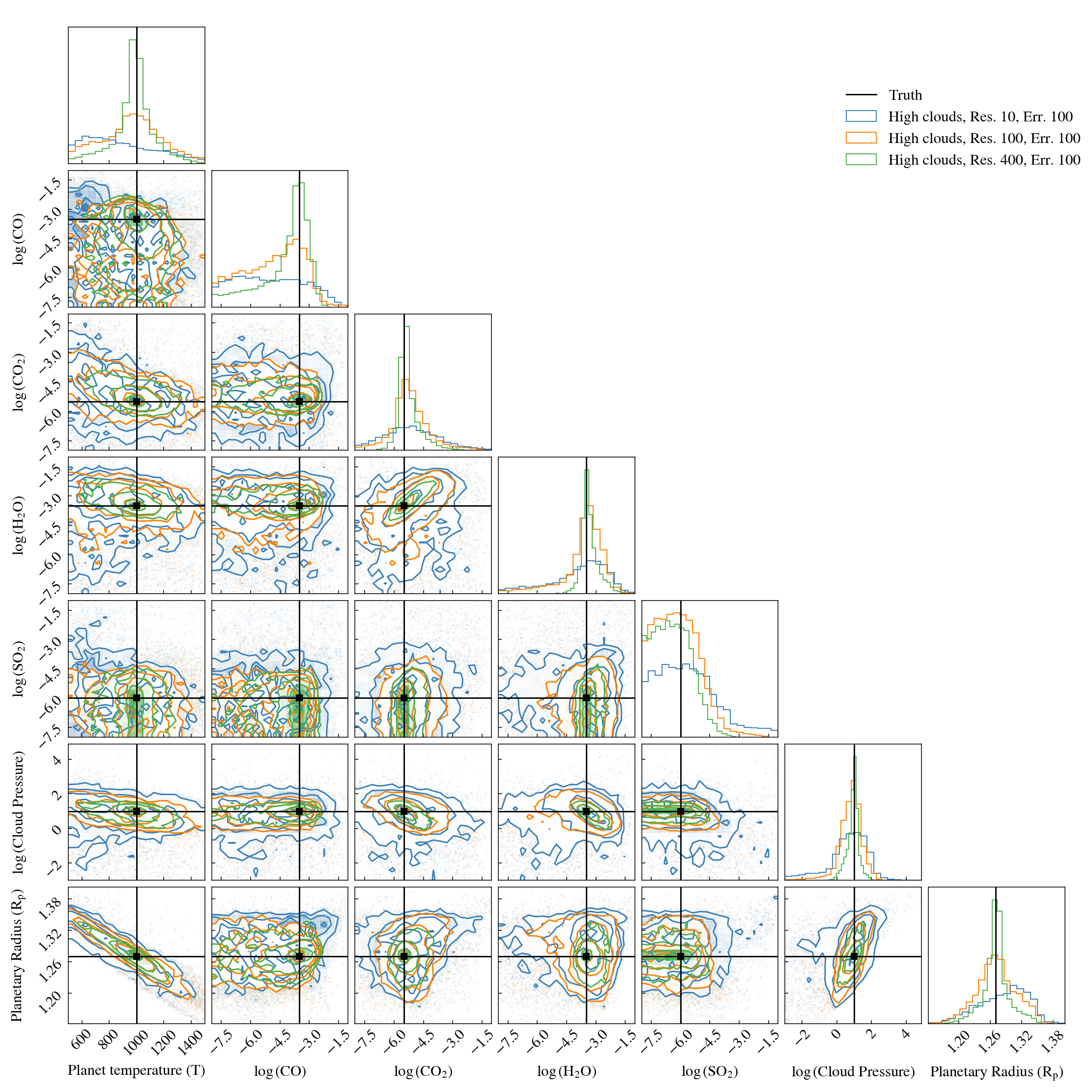}
    \caption{A corner plot of the retrievals used to produce the spectrum contribution plots in figure~\ref{fig:SpecContributions}. The three different colours represent retrievals on spectra at three different resolutions. }
    \label{fig:CornerPlot_ResolveSO2}
\end{figure*}

Here we include corner plots for the same retrievals as those used to produce the contribution plots shown in the main text, figure~\ref{fig:SpecContributions}. Presented in figure~\ref{fig:CornerPlot_ResolveSO2}, these retrievals use high cloud simulations at resolutions of 10, 100 and 400. They are plotted on the same set of corner plot axes in order to demonstrate how we are better able to constrain most of the parameters with increasing resolution but the log-abundance of SO\textsubscript{2} remains poorly retrieved since its features are masked by the cloud deck.

\subsection{Binning from ERS grid}
\label{App:BinFactor}

\begin{figure}
    \includegraphics[width=\columnwidth]{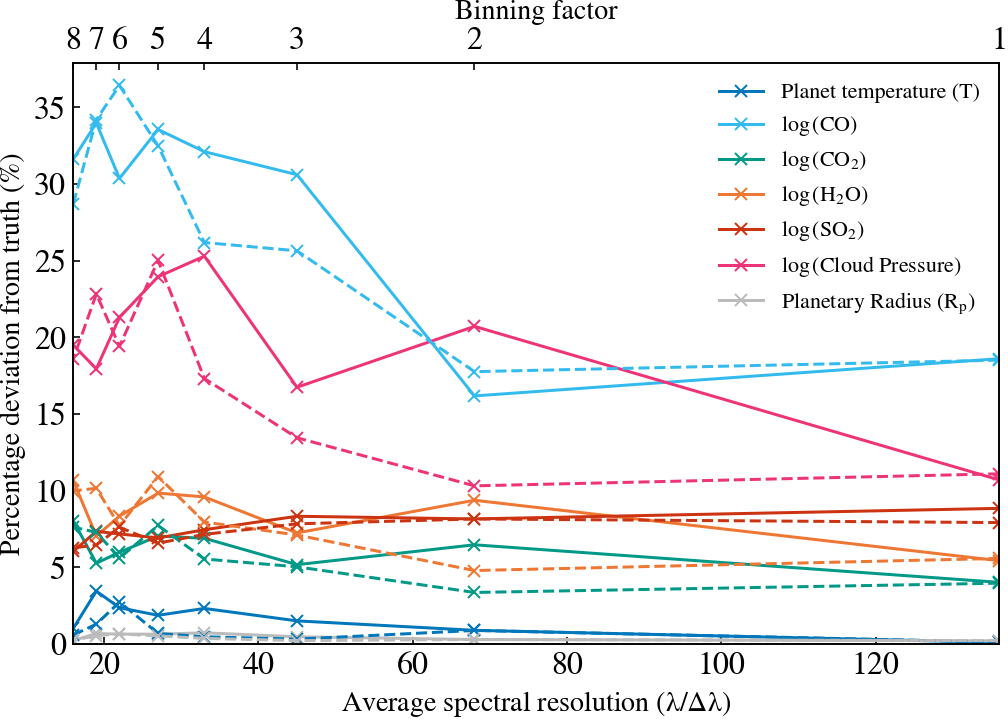}
    \caption{The percentage deviation of retrieved parameters from spectra obtained by binning down simulations of the JWST ERS data for WASP-39b using the NIRSpec PRISM \citep{NIRSpecPRISM_Wasp39b} with a high cloud deck. The wavelength grid was constructed by retaining only every N\textsuperscript{th} point in the spectrum where N is the 'Binning factor'. Each line represents a different parameter and the top and bottom x-axes quantify the level of binning through the binning factor and the average resolution  respectively. All values for average resolution are obtained by considering the mean value across the spectrum. Solid lines show data from retrievals where the simulation was first binned to the wavelength grid of the ERS data and then binned further whereas the dashed lines bin directly from the native resolution of the TauREx simulation to the appropriate reduced grid.}
    \label{fig:HighCloudBinFac}
\end{figure}

\begin{figure}
    \includegraphics[width=\columnwidth]{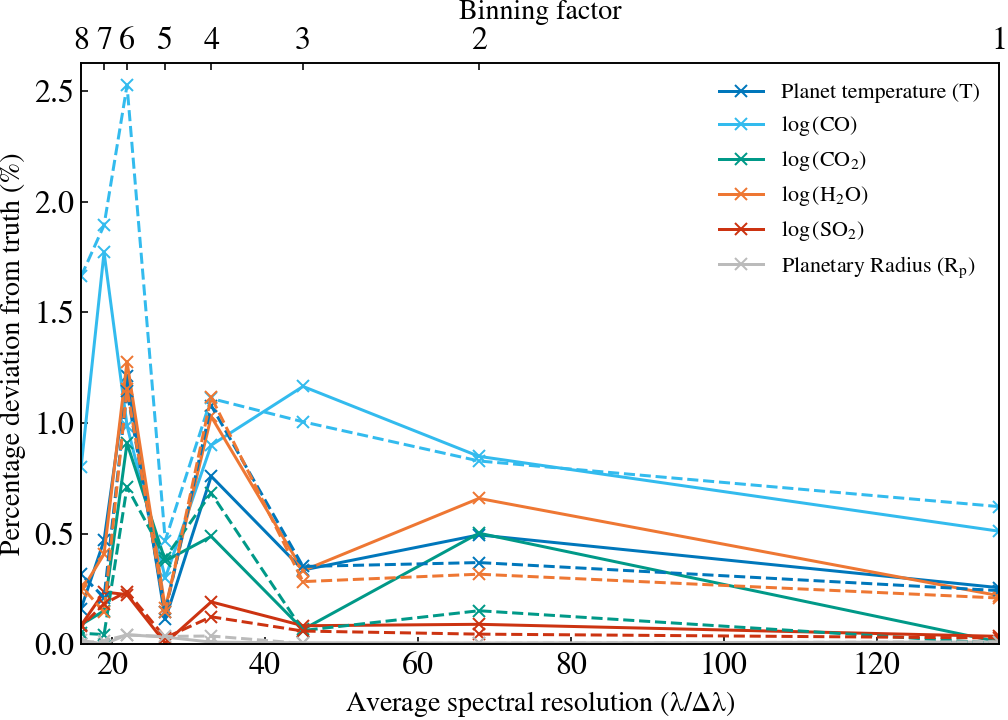}
    \caption{The same as figure~\ref{fig:HighCloudBinFac} except that these retrievals use cloud free spectra. Note that the values for cloud pressure have been omitted since they cannot be accurately retrieved in this set up and therefore their large deviations would make the variation in other parameters difficult to view.}
    \label{fig:NoCloudBinFac}
\end{figure}

Using the same grids as discussed in section~\ref{sec:BinFactorRes}, we run retrievals on the simulations for the high cloud and no cloud cases (with results shown in figures~\ref{fig:HighCloudBinFac} and \ref{fig:NoCloudBinFac} respectively). In both cases we observe similar trends when retrieving parameters from simulations binned directly to the reduced grid (dashed lines) and when binning first to the ERS resolution (solid lines).

\subsection{Binning from ERS Grid with a shifted grid}
\label{App:BinFactor_shift}

As in the previous section, plots in this section use retrievals on spectra which have been reduced using the methods discussed in section~\ref{sec:BinFactorRes}. However, in this case some of the wavelength grids receive an artificial shift before the spectra are binned. These shifts are set such that the wavelength points in the shifted grid will fall halfway between the corresponding point and the next point in the original wavelength grid.

\begin{figure}
    \includegraphics[width=\columnwidth]{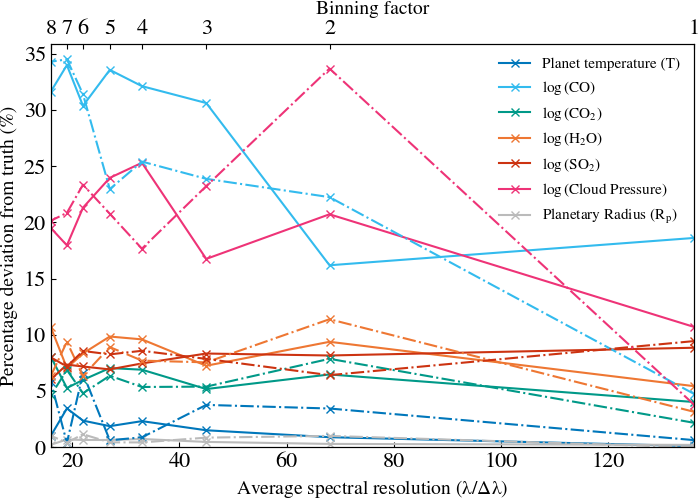}
    \caption{The percentage deviation of retrieved parameters from spectra obtained by binning down simulations of the JWST ERS data for WASP-39b using the NIRSpec PRISM \citep{NIRSpecPRISM_Wasp39b} with a high cloud deck. The wavelength grid was constructed by retaining only every N\textsuperscript{th} point in the spectrum where N is the 'Binning factor'. Each line represents a different parameter and the top and bottom x-axes quantify the level of binning through the binning factor and the average resolution  respectively. All values for average resolution are obtained by considering the mean value across the spectrum. Solid lines show data from retrievals where the simulation was first binned to the wavelength grid of the ERS data and then binned further. These lines have a reduced opacity in order to highlight the new data shown by the dot-dashed lines. These results also have a shift applied to the wavelength grid prior to binning the simulation.}
    \label{fig:HighCloudBinFac_shift}
\end{figure}

\begin{figure}
    \includegraphics[width=\columnwidth]{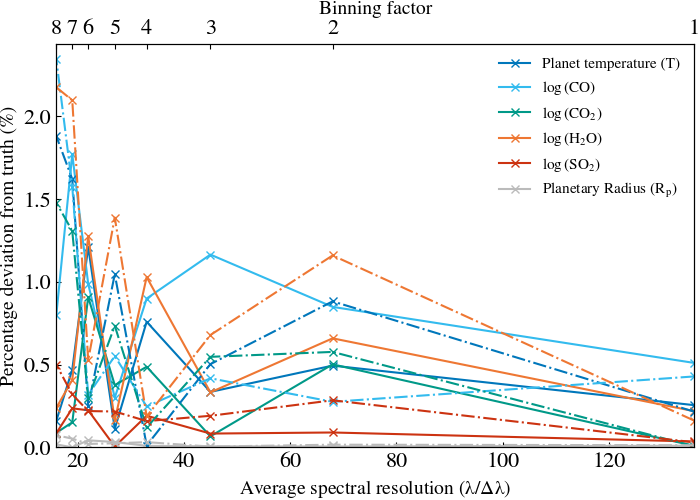}
    \caption{The same as figure~\ref{fig:HighCloudBinFac_shift} except that these retrievals use cloud free spectra. Note that the values for cloud pressure have been omitted since they cannot be accurately retrieved in this set up and therefore their large deviations would make the variation in other parameters difficult to view.}
    \label{fig:NoCloudBinFac_shift}
\end{figure}

Plots compare the percentage deviation (the accuracy) of the retrieved parameters when retrieving from spectra that have been binned to grids with and without the additional shift applied. The plot for data using a low cloud simulation is included in the main text (figure~\ref{fig:BinFactorDev_withShift}) and the results for the high and no cloud case are included here (figures~\ref{fig:HighCloudBinFac_shift} and \ref{fig:NoCloudBinFac_shift} respectively).

\subsection{Reduced model and retrieval complexity}
\label{App:Complexity}

\begin{figure*}
    \includegraphics[width=\textwidth]{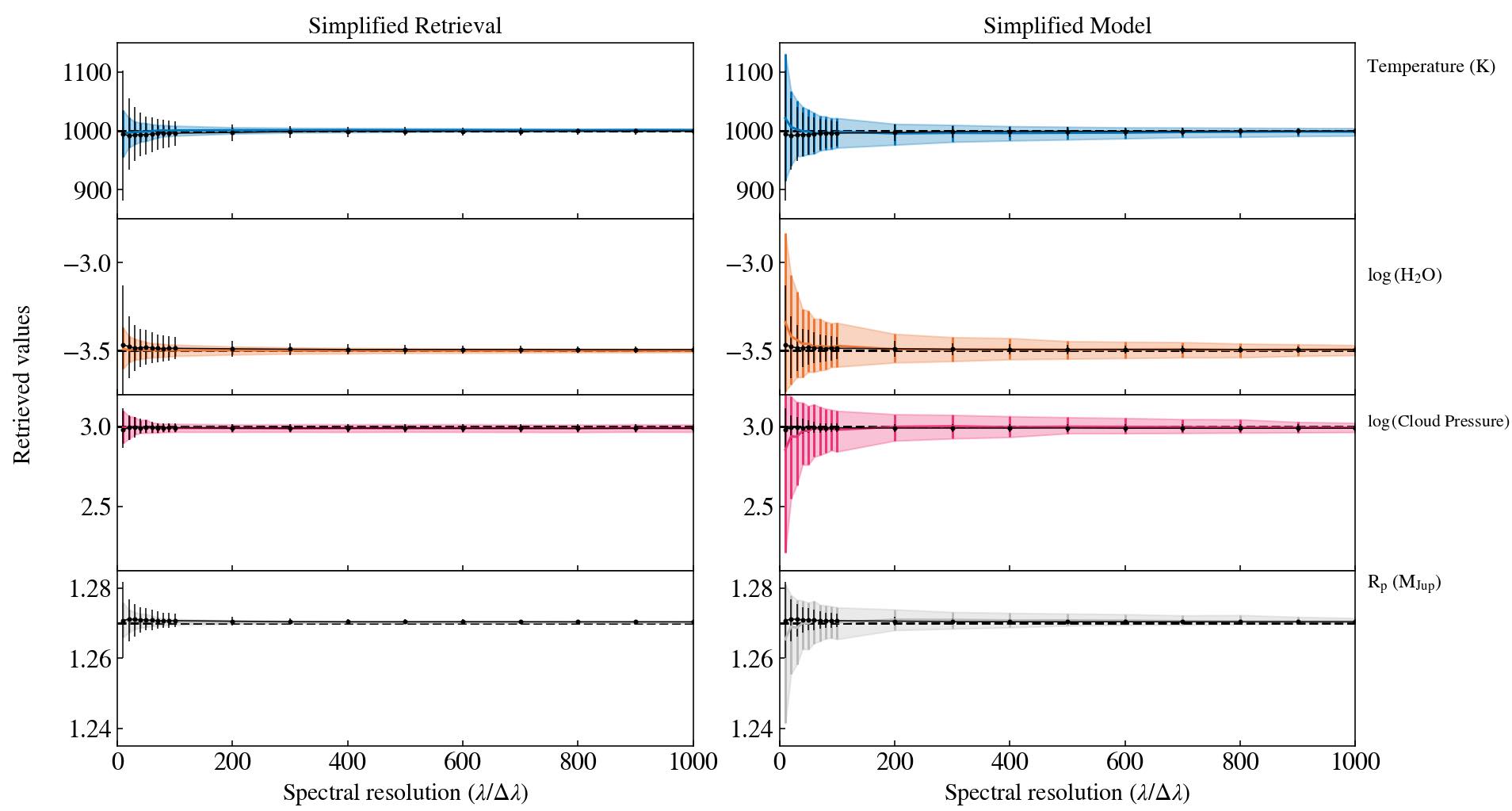}
    \caption{Comparing the retrieved values for parameters in reduced complexity models and retrievals. On both the left and the right columns, data in black show the results from the original resolution-error grid. Then, on the left, coloured data represent the retrieved values when a spectrum containing H\textsubscript{2}O as the only atmospheric molecule is input. In the right column, coloured data represent retrieved values when a spectrum featuring the same complexity as the original model is used but the retrieval only fits for the molecular log-abundance of H\textsubscript{2}O. All of the other molecular abundances are fixed to their correct values. All of these simulations used the low cloud scenario and errors of 100\,ppm.}
    \label{fig:OnlyH2OFit}
\end{figure*}

In our investigation of changing model complexity, we compare the results in the case of the original retrievals and only fitting for certain molecules in the retrievals as well as cases where a model of reduced complexity is used in the first place. Figure~\ref{fig:OnlyH2OFit} shows results in the cases where we use a spectrum only containing water and a spectrum containing all of the same molecules as our initial investigation but only fitting for H\textsubscript{2}O in the retrieval. In this second case, other molecules have their log-abundances fixed to their true values. These two tests use the low cloud case and constant errors bars of 100\,ppm (since these points in our grid probe near to  the average error of the JWST ERS data).  

\subsection{Repeat retrieval instances}
\label{App:Repeat}

\begin{figure}
    \includegraphics[width=\columnwidth]{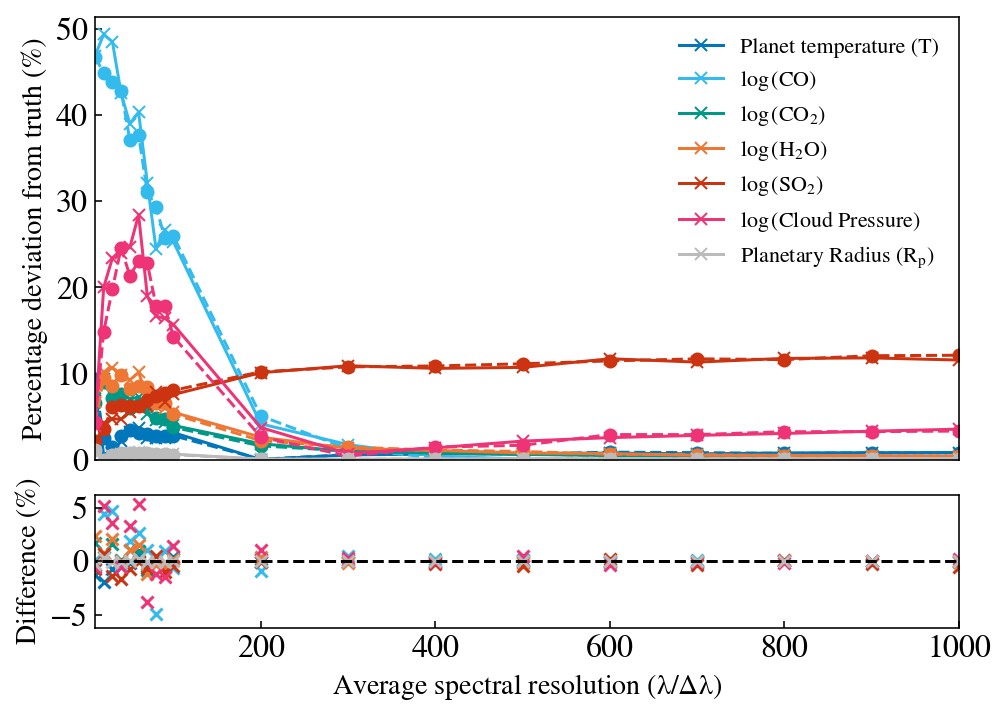}
    \caption{Two set of retrieval results for identical input spectra (high cloud simulations with errors of 100\,ppm). The top panel plots the two sets of results against one another with the original data using a solid line and crosses and the repeat instance of the retrievals using a dashed line and circles. Parameter are separated by colours in the plot. The lower panel shows the difference in the percentage deviation for each parameter for each retrieval.}
    \label{fig:HighCloudRep}
\end{figure}

\begin{figure}
    \includegraphics[width=\columnwidth]{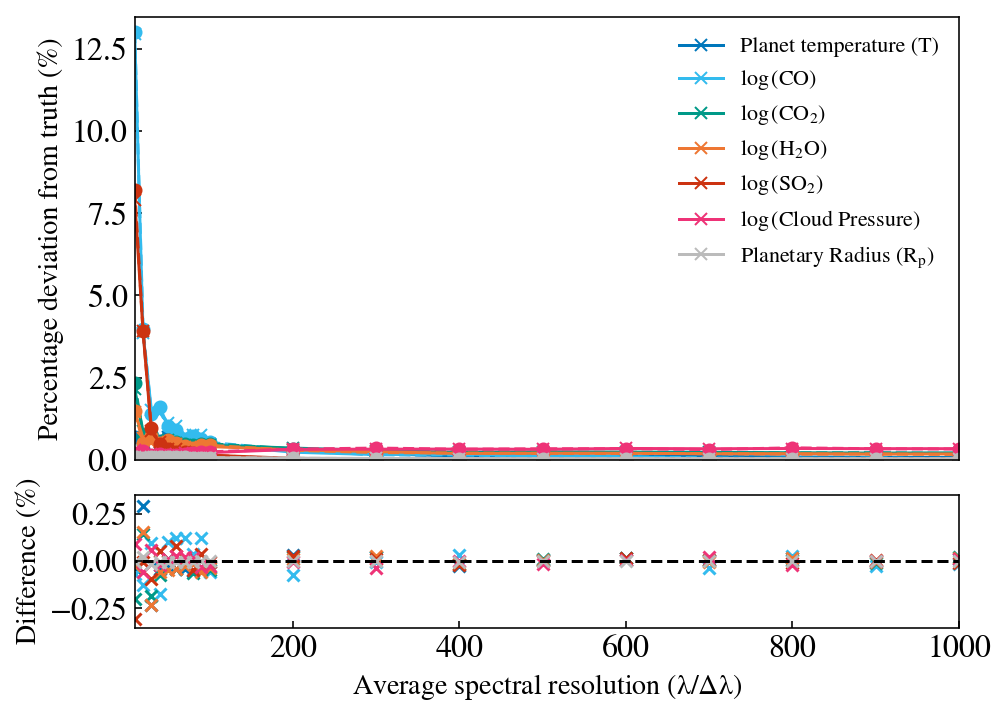}
    \caption{The same as figure~\ref{fig:HighCloudRep} except that these retrievals use low cloud spectra.}
    \label{fig:LowCloudRep}
\end{figure}

\begin{figure}
    \includegraphics[width=\columnwidth]{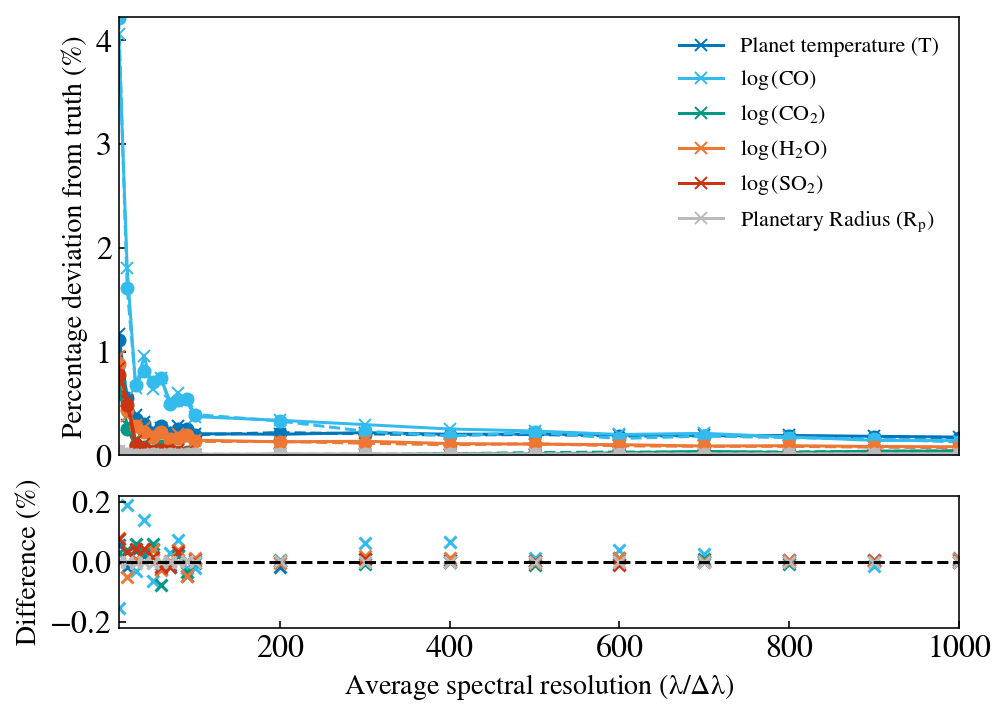}
    \caption{The same as figure~\ref{fig:HighCloudRep} except that these retrievals use cloud free spectra.}
    \label{fig:NoCloudRep}
\end{figure}

Plots in this section consider the results from repeated retrievals. By this we mean cases where retrievals were re-run on spectra with constant resolution grids and constant error bars (set at 100\,ppm). We include plots for the high, low and no cloud simulations (figures~\ref{fig:HighCloudRep}, \ref{fig:LowCloudRep} and \ref{fig:NoCloudRep} respectively). Overall we do not see any significant differences between the original sets of retrievals and these repeat instances. Our retrievals return the same predictions in both cases (within error) indicating the stability of these results despite the stochatsic sampling processes used. 

\subsection{Retrievals on scattered spectra}
\label{App:Perturb}

To investigate the effect of adding additional noise to the data, we scatter a subset of the spectra from our original retrieval grid. The spectra used are low cloud simulations with errors of 50, 100 and 200\,ppm.

\begin{figure*}
    \includegraphics[width=\textwidth]{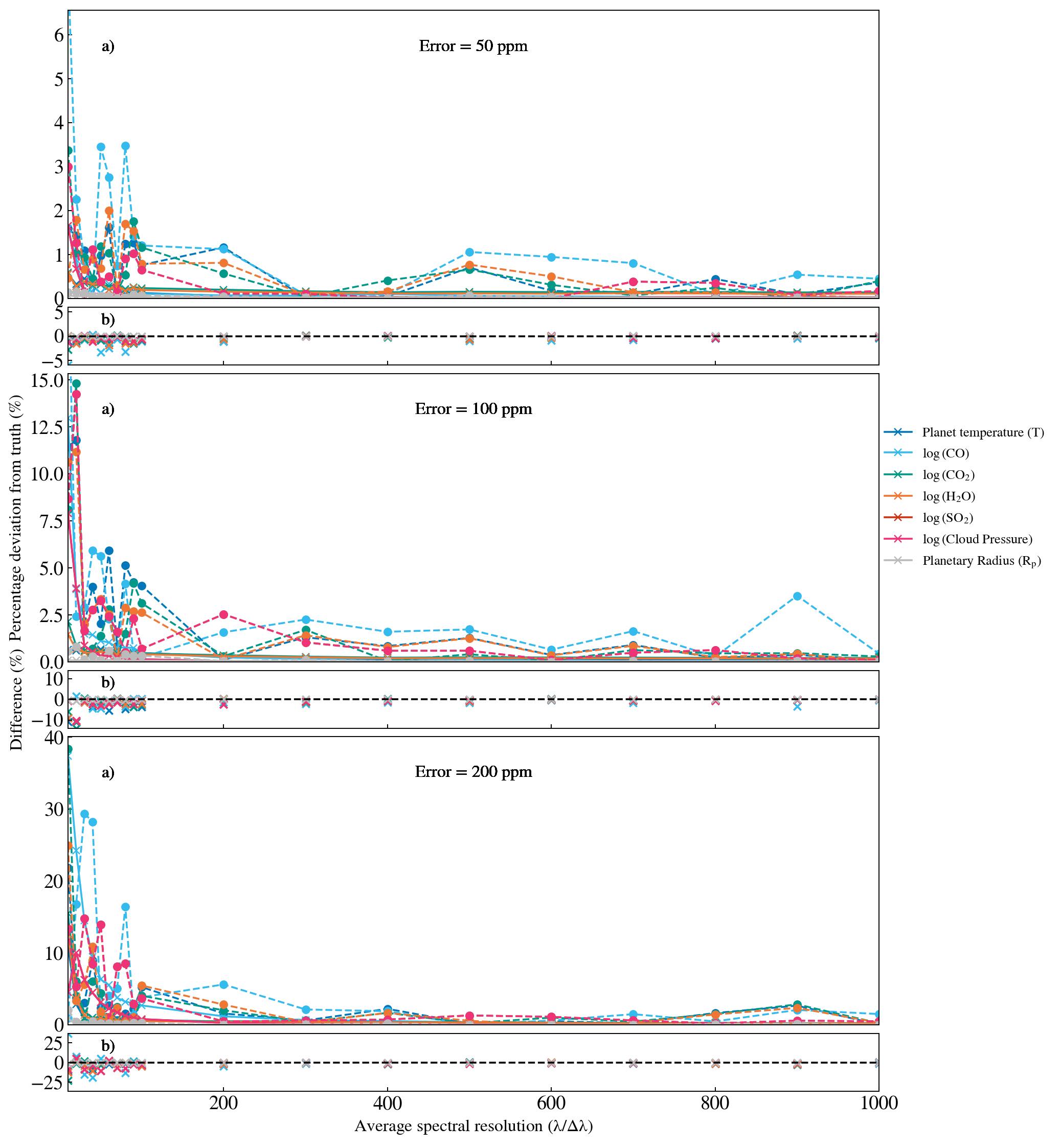}
    \caption{Comparison of two set of retrieval results for simulations of a low cloud spectrum with errors of 50, 100 or 200\,ppm. In each pair of plots (an \textbf{a)} and a \textbf{b)} plot), one retrieval is run on a spectrum without any noise added and one is run on a spectrum where the data have been scattered according to a Gaussian distribution centred on the spectrum point and with a width set by the size of the error bar. The \textbf{a)} panels plot the two sets of results against one another with the non-scattered data using a solid line and crosses and the perturbed instance of the retrievals using a dashed line and circles. Parameter are separated by colours in their plots. The \textbf{b)} panels show the difference in the percentage deviation for each parameter for each retrieval between the scattered and non-scattered datasets.}
    \label{fig:LowCloudScatter}
\end{figure*}

With dashed lines and circular markers, we plot the new data in the \textbf{a)} panels of figure~\ref{fig:LowCloudScatter} alongside the corresponding retrievals from the original grid plotted with solid lines and cross markers. Then, in the \textbf{b)} panels, we plot the differences in the percentage deviation of each parameters prediction when retrieving from a scattered or noise-free simulation.

Since we are only using one instance of the random resampling of spectra, we expect there to be some outliers in the predictions due to particular noise instances. However, these plots help to show the extent of the possible error emerging from one specific noise instance. With a photometric error of 200\,ppm on every spectral point, the percentage deviations can differ by more than 25\,\% between the noise-free and noisy spectra. However, it is only about 5\,\% for the worst case in the the spectra with errors of 50\,ppm. Additionally, it is worth noting that these significant differences are seen at the low resolution ends of our tested grids since here the data is more dependent on its accuracy in transit depth than at the high resolution end where this is accounted for by finer wavelength coverage. 

%%%%%%%%%%%%%%%%%%%%%%%%%%%%%%%%%%%%%%%%%%%%%%%%%%

% Don't change these lines
\bsp	% typesetting comment
\label{lastpage}
\end{document}